\documentclass[twocolumn]{aa}
\usepackage{graphicx}
\begin{document}

\authorrunning{Nagao et al.}
\titlerunning{Gas metallicity diagnostics}

\title{Gas metallicity diagnostics in star-forming galaxies}

\author{
          Tohru Nagao            \inst{1, 2},
          Roberto Maiolino       \inst{1, 3}, \and
          Alessandro Marconi     \inst{1}
}

\offprints{T. Nagao}

\institute{
           INAF -- Osservatorio Astrofisico di Arcetri,
           Largo Enrico Fermi 5, 50125 Firenze, Italy\\
           \email{tohru@arcetri.astro.it,
                  maiolino@arcetri.astro.it,
                  marconi@arcetri.astro.it}
           \and
           National Astronomical Observatory of Japan,
           2-21-1 Osawa, Mitaka, Tokyo 151-8588, Japan
           \and
           INAF -- Osservatorio Astrofisico di Roma,
           Via di Frascati 33, 00040 Monte Porzio Catone, Italy
          }

\date{Received ;  accepted }

\abstract{
  Generally the gas metallicity in distant galaxies can only 
  be inferred by using a few prominent emission lines. 
  Various theoretical models have been used to predict
  the relationship between emission line fluxes and metallicity,
  suggesting that some line ratios can be used as diagnostics of 
  the gas metallicity in galaxies. However, accurate empirical 
  calibrations of these emission line flux ratios from real 
  galaxy spectra spanning a wide metallicity range are still 
  lacking. In this paper we provide such empirical calibrations 
  by using the combination of two sets of spectroscopic data: 
  one consisting of low-metallicity galaxies with a 
  measurement of [O{\sc iii}]$\lambda$4363 taken from the 
  literature, including spectra from
  the Sloan Digital Sky Survey (SDSS), and the other one 
  consisting of galaxies in the SDSS database whose gas 
  metallicity has been determined from various strong emission 
  lines in their spectra. This combined data 
  set constitutes the largest sample of galaxies with 
  information on the gas metallicity available so far and 
  spanning the widest metallicity range. By using these data we 
  obtain accurate empirical relations between gas metallicity 
  and several emission line diagnostics, including the $R_{23}$ 
  parameter, the [N{\sc ii}]$\lambda$6584/H$\alpha$ and
  [O{\sc iii}]$\lambda$5007/[N{\sc ii}]$\lambda$6584 ratios. 
  Our empirical diagrams show that the line ratio 
  [O{\sc iii}]$\lambda$5007/[O{\sc ii}]$\lambda$3727 is a 
  useful tool to break the degeneracy in the $R_{23}$ parameter 
  when no information on the [N{\sc ii}]$\lambda$6584 line is 
  available. The line ratio 
  [Ne{\sc iii}]$\lambda$3869/[O{\sc ii}]$\lambda$3727 also 
  results to be a useful metallicity indicator for high-$z$ 
  galaxies, especially when the $R_{23}$ parameter or other 
  diagnostics involving [O{\sc iii}]$\lambda$5007 or 
  [N{\sc ii}]$\lambda$6584 are not available. 
  Additional, useful diagnostics newly proposed in this
  paper are the line ratios of
  (H$\alpha$+[N{\sc ii}]$\lambda\lambda$6548,6584)/[S{\sc ii}]$\lambda$6720,
  [O{\sc iii}]$\lambda$5007/H$\beta$, and
  [O{\sc ii}]$\lambda$3727/H$\beta$. Finally, we 
  compare these empirical relations with photoionization 
  models. We find that the empirical $R_{23}$-metallicity 
  sequence is strongly discrepant with respect to the trend 
  expected by models with constant ionization parameter. 
  Such a discrepancy is also found for other line ratios. These
  discrepancies provide evidence for a strong 
  metallicity dependence of the average ionization parameter
  in galaxies. In particular, we find that the average 
  ionization parameter in galaxies increases by $\sim 0.7$~dex 
  as the metallicity decreases from 2~Z$_{\odot}$ to 
  0.05~Z$_{\odot}$, with a small dispersion. This result should 
  warn about the use of theoretical models with constant 
  ionization parameter to infer metallicities from observed 
  line ratios.
    \keywords{
              galaxies: abundances  --
              galaxies: evolution   --
              galaxies: general     --
              galaxies: ISM         --
              H{\sc ii} regions
             }
}

\maketitle

\section{Introduction}

The gas metallicity is one of the most important tools to 
investigate the evolutionary history of galaxies. This is because 
the gas metallicity of galaxies is basically determined by their
star-formation history. Recent observational studies allowed the
investigation of the gas metallicity even in high-$z$ galaxies
beyond $z = 1$, such as Lyman-break galaxies (e.g., 
Teplitz et al. 2000a, 2000b; Pettini et al. 2001),
submillimeter-selected high-$z$ galaxies (Swinbank et al. 2004),
and so on (see also, e.g., Kobulnicky \& Kewley 2004;
Savaglio et al. 2005; Maier et al. 2006; Liang et al. 2006;
Erb et al. 2006).
Such observational insights on the metallicity evolution of 
galaxies are now giving constraints on the theoretical 
understandings of the formation and the evolution of galaxies 
(e.g., Bicker et al. 2004).

However, metallicity measurements in distant galaxies are not
straightforward. Information on the gas temperature is required 
for a precise  determination of the gas metallicity, but the gas 
temperature can be accurately inferred only when the fluxes of 
auroral emission lines such as [O{\sc iii}]$\lambda$4363 and 
[N{\sc ii}]$\lambda$5755 are known, and these are generally too 
weak to be measured in faint distant galaxies. The measurement of 
the auroral emission lines is difficult even for galaxies in the 
local universe especially when the gas metallicity is high, 
because the collisional excitation of the auroral transitions is 
suppressed due to efficient cooling through far-infrared 
fine-structure emission lines (e.g., Ferland et al. 1984;
Nagao et al. 2006a). Therefore, in general we have to rely on 
some relations between gas metallicity and flux ratios of strong 
emission-lines to estimate the gas metallicity in most galaxies. 
Extensive studies have been performed to calibrate such 
metallicity diagnostics by using only strong emission lines. One of 
the most frequently used metallicity diagnostics is the $R_{23}$ 
parameter, defined as
\begin{equation}  % equation (1)
  R_{23} \! = \! 
         \frac{ F({\rm [OII]}\lambda3727) \!\! + \!\!
                F({\rm [OIII]}\lambda4959) \!\! + \!\!
                F({\rm [OIII]}\lambda5007) } 
              { F({\rm H}\beta\lambda4861) },
\end{equation}
where $F$([O{\sc ii}]$\lambda$3727), 
$F$([O{\sc iii}]$\lambda$4959) and so on denote the 
emission-line fluxes of [O{\sc ii}]$\lambda$3727, 
[O{\sc iii}]$\lambda$4959 and so on, respectively. 
% Note that $F$([S{\sc ii}]$\lambda$6720) denotes the sum of 
% $F$([S{\sc ii}]$\lambda$6717) and $F$([S{\sc ii}]$\lambda$6731).
The $R_{23}$ was proposed by Pagel et al. (1979), and its 
calibration to the oxygen abundance has been improved by various 
photoionization model calculations (e.g., McGaugh 1991; 
Kewley \& Dopita 2002).

One serious problem of this indicator is that a certain value 
of $R_{23}$ has two different solutions, a low-metallicity 
solution and a high metallicity one. Therefore additional, or 
alternative, diagnostics aimed at removing the $R_{23}$ 
degeneracy have been proposed (e.g., Alloin et al. 1979; 
Denicol\'{o} et al. 2002; Kewley \& Dopita 2002; 
Pettini \& Pagel 2004). However, most of these methods exploit 
the [N{\sc ii}]$\lambda$6584 line, which has the problem of 
being very weak at sub-solar metallicities (hence difficult to 
measure) and the problem of being rapidly shifted outside the 
spectral band of many surveys at high redshift (e.g., unusable 
beyond $z \sim 0.5$ in optical spectra). On the theoretical 
side various models have been presented, which provide the 
ratios among the most prominent emission lines as a function 
of metallicity (e.g. Kewley and Dopita 2002). However, model
predictions strongly depend on the assumed physical parameters 
of the ionized gas, and in particular on the ionization 
parameter [$U \equiv \Phi_{\rm H} / (c n_{\rm H})$, where
$\Phi_{\rm H}$ is the surface flux of hydrogen-ionizing 
photon and $n_{\rm H}$ is the gas density]. As a consequence, 
an accurate correspondence between individual diagnostics 
(line ratios) and metallicity cannot be established, because 
of the lack of information on the physical conditions of the 
gas. Summarizing, many gas metallicity diagnostics proposed 
so far are either ambiguous or unusable when applied to the 
spectra of distant galaxies. 

The goal of this paper is to obtain accurate, empirical 
calibrations between metallicity and individual diagnostics 
involving a few strong emission lines, which can be applied 
to the spectra of distant galaxies. In particular, we 
re-calibrate diagnostics already proposed in the past, but 
we also propose new diagnostics which appear particularly 
suited for distant galaxies. This work is obtained by 
combining two large data sets. The first one is composed of 
recent spectroscopic observations of low-metallicity galaxies 
[7.0$\la$12+log(O/H)$\la$8.5], whose metallicity is accurately 
determined through the [O{\sc iii}]$\lambda$4363 line (\S\S2.1). 
This dataset consists of two subsamples; one is
taken from the database of the Sloan Digital Sky Survey (SDSS; 
York et al. 2000; Strauss et al. 2002) (\S\S2.1.1) and the 
other is taken from the literature (\S\S2.1.2).
The second data set is a subsample of galaxies in the SDSS
database, whose metallicity has been derived by Tremonti et al. 
(2004) [8.2$\la$12+log(O/H)$\la$9.2; see \S\S2.2]. These 
combined data sets provide the largest sample of galaxies with 
information on the gas metallicities and spanning more than 
2 dex in metallicity.

\section{Data}

\subsection{Spectroscopic data of low-metallicity galaxies
            with a [OIII]$\lambda$4363 measurement}

\subsubsection{SDSS data (sample A)}

The gas-phase oxygen abundance is well determined when
the flux of [O{\sc iii}]$\lambda$4363 is measured (e.g.,
Osterbrock 1989). Although such measurements have been
performed for more than a hundred low-metallicity galaxies,
simple compilation of those earlier results may introduce 
some unexpected biases and uncertainties. This is because
the data were collected by various (heterogeneous)
observations with different properties (aperture size,
wavelength resolution, and so on) and because the method of
calculating the oxygen abundance is different for different
authors. Recently, Izotov et al. (2006b) reported their 
systematic measurements of the oxygen abundance for
low-metallicity galaxies in the SDSS Data Release 3 (DR3;
Abazajian et al. 2005) by using the [O{\sc iii}]$\lambda$4363
emission-line flux. The extinction-corrected emission-line
fluxes of galaxies with a measurement of the oxygen abundance
provided by Izotov et al. (2006b) are the ideal data for the
empirical calibration of metallicity diagnostics, because
the data were obtained and measured in a homogeneous way and
because the oxygen abundance is also calculated with a
common method. The number of spectra analyzed by Izotov et al.
(2006b) is 309. Among them, we use the data with a relatively
small error in the oxygen abundance 
[$\Delta$(log$\frac{\rm O}{\rm H}) \leq$ 0.05] 
(146 spectra).
Here we adopt the uncertainty 
[$\Delta$(log$\frac{\rm O}{\rm H}$)] given in Table 2 of 
Izotov et al. (2006).
Since some spectra in the database of Izotov et al. (2006b)
are duplicated for the same objects, the number of galaxies
with $\Delta$(log$\frac{\rm O}{\rm H}) \leq$ 0.05 is 
139. Hereafter we call this sample ``sample A''.

However, this sample has two problems when used to accurately
calibrate metallicity diagnostics. First, most of galaxies in 
sample A have a relatively high oxygen abundance, and only 6 of
them have 12+log(O/H) $<$ 7.6. Therefore the 
statistical reliability
of the empirical calibration of metallicity diagnostics would
be extremely poor at 12+log(O/H) $<$ 7.6
if using only this sample. Second, in sample A,
there is the remarkable tendency for lower-metallicity galaxies
to have lower redshift. In Figure 1, the oxygen abundance of 
galaxies in sample A is shown as a function of redshift.
The origin of this apparent tendency is likely due to the fact
that SDSS is not a volume-limited 
survey; that is, galaxies at higher redshift have 
preferentially higher luminosity, and thus higher metallicity.
In particular, all of the galaxies with 12+log(O/H) $<$ 7.6
are at $z < 0.02$. This means that the [O{\sc ii}]$\lambda$3727
flux cannot be measured for the latter galaxies due to the limited
wavelength coverage of the SDSS spectroscopy 
($\lambda_{\rm obs} \ga 3800{\rm \AA}$). Therefore, if using
only sample A, we could not calibrate the diagnostics involving
[O{\sc ii}]$\lambda$3727 [$R_{23}$ and 
$F$([N{\sc ii}]$\lambda$6584)/$F$([O{\sc ii}]$\lambda$3727)]
in the metallicity range 12+log(O/H) $<$ 7.6. This is a 
serious problem,
because $R_{23}$ is one of the most frequently used metallicity
diagnostics and thus should be calibrated in a wide metallicity
range. In conclusion, the accurate calibration of various
metallicity diagnostics in a wide metallicity range cannot be
achieved by using only sample A. We therefore collected
additional data of [O{\sc iii}]$\lambda$4363-detected galaxies, 
which are described in the following
subsection.

\subsubsection{Other data from literature (sample B)}

In order to increase the number of low-metallicity galaxies
with a measurement of the oxygen abundance, we compiled the
reddening-corrected emission-line flux data of galaxies
with a [O{\sc iii}]$\lambda$4363 measurement from the literature.
The sample of compiled galaxies is given in Table 1.
For the objects whose spectroscopic 
properties have been reported by more than one paper 
independently, we chose the one with higher signal-to-noise ratio.
When both the spectroscopic properties of the whole galaxy 
and of parts of it have been reported, we compiled both of 
them (e.g., Mrk 116). Consequently, the number of the 
compiled objects is 157. To minimize possible systematic 
errors owing to the different methods on the calculation of 
the oxygen abundance, we re-calculate their oxygen abundance 
by adopting the same method used for sample A
(Izotov et al. 2006b).
The re-calculated $R_{23}$ parameter, gas density of
[S{\sc ii}]-emitting region [$n_{\rm H}$(S$^+$)], gas
temperature of [O{\sc iii}]-emitting region 
[$t{\rm (O^{2+})}$] and oxygen abundance [$12 + $log(O/H)] 
are given in Table 1, 
along with the reference to the data of the emission-line 
flux ratios. To calculate $R_{23}$, we did not use
$F$([O{\sc iii}]$\lambda$4959) but calculate the ratio of
[$F$([O{\sc ii}]$\lambda$3727) + 
1.327$\times$$F$([O{\sc iii}]$\lambda$5007)]
$\! \! / \! $$F$(H$\beta$) since only the flux of 
[O{\sc iii}]$\lambda$5007 (without that of 
[O{\sc iii}]$\lambda$4959) is given in some reference papers. 

To check whether our adopted method causes possible systematic 
difference in the oxygen abundance from the values given in the 
original references, we compare the oxygen abundances 
re-calculated by us and those given in the original papers in 
Figure 3. Apparently, there is no systematic difference between 
our results and the results given in the literature. The mean 
and the RMS of the difference, [12+log(O/H)$_{\rm this \ work}$] 
-- [12+log(O/H)$_{\rm literature}$], are +0.001 and 0.041,
respectively. This 
mean value of the difference is smaller than the typical error 
of the re-calculated oxygen abundance.
Among the 157 objects given in Table 1 and plotted in Figure 3, 
we use only data
with $\Delta$(log$\frac{\rm O}{\rm H}) \leq$ 0.05. 
This constraint results in a sample of 120 objects, that is 
hereafter called ``sample B''.
Note that the mean and the RMS of the difference, 
[12+log(O/H)$_{\rm this \ work}$] -- 
[12+log(O/H)$_{\rm literature}$], for sample B are
--0.008 and 0.038. Again the mean value of the difference is 
smaller than the typical error on the oxygen abundance.

\subsection{Spectroscopic data of high-metallicity galaxies
            (sample C)}

For high-metallicity galaxies, we referred to the oxygen 
abundance derived by Tremonti et al. (2004), who derived the 
metallicities of $\sim$53000 galaxies in the SDSS database.
They used the fluxes of many strong emission lines 
([O{\sc ii}]$\lambda$3727, H$\beta \lambda$4861, 
[O{\sc iii}]$\lambda$5007, H$\alpha$, [N{\sc ii}]$\lambda$6584, 
[S{\sc ii}]$\lambda$6717, and [S{\sc ii}]$\lambda$6731) and 
comparing them with  photoionization models
(Ferland et al. 1998). Although they 
presented the results of their analysis on the spectra of the 
SDSS Data Release 2 (DR2; Abazajian et al. 2004), they also 
provide the results of their recent analysis on the spectra 
from Data Release 4 (DR4; Adelman-McCarthy et al. 2006) on 
their web site\footnote{http://www.mpa-garching.mpg.de/SDSS/}. 
Their estimate of the oxygen abundance 
does not rely only on a single metallicity diagnostic flux 
ratio, but uses all the optical prominent emission lines (see 
also, Charlot \& Longhetti 2001; Brinchmann et al. 2004).
Therefore, among galaxies 
without [O{\sc iii}]$\lambda$4363 flux, their sample is 
currently the best one in terms of both sample size and 
reliability.

The oxygen-abundance catalog of the SDSS DR4 galaxies 
contains 567486 objects. The objects in the catalog are 
classified into the five classes; ``star-forming galaxies'', 
``low S/N star-forming galaxies'', ``composite'', ``active 
galactic nuclei (AGNs)'', and ``unclassificable''. We 
referred only to the galaxies belonging the first class 
(141317 objects). 
The emission-line fluxes of these galaxies are obtained from 
the emission-line flux catalog of the SDSS DR4 galaxies
provided on the same web site as the oxygen-abundance catalog.
The emission-line fluxes given in this catalog were measured
from the stellar-continuum subtracted spectra with the latest 
high spectral resolution population synthesis models by 
Bruzual \& Charlot (2003), and thus more reliable than the 
flux data provided on the SDSS Data Archive Server. Since the 
emission-line fluxes given in their catalog are not corrected 
for dust extinction, we corrected them by using the Balmer 
decrement method with the reddening curve of Cardelli et al. 
(1989). We then removed the duplicated objects and the objects 
observed in some problematic plates (see the SDSS web 
page\footnote{http://www.sdss.org/dr4/}) from the cross sample 
of the oxygen-abundance catalog and the emission-line flux 
catalog. Then we select only objects satisfying all of the 
following five criteria:
\begin{enumerate}
  \item The redshift is higher than 0.028.
  \item Both H$\alpha$ and H$\beta$ emission lines have 
        S/N $\geq$ 10.
  \item log$\frac{F({\rm [OIII]\lambda5007})}{F({\rm H\beta})}$ 
        $>$ $\frac{0.61}{{\rm log}\frac{F({\rm [NII]\lambda6584})}
        {F({\rm H\alpha})}-0.05}$ + 1.3.
  \item The fiber aperture covers at least 20\% of the total 
        $g^\prime$-band photons.
  \item The uncertainty on the estimated stellar mass is less 
        than 0.5 dex (i.e., log$M_{97.5}$ -- log$M_{2.5}$ $<$ 
        0.5, where $M_{97.5}$ and $M_{2.5}$ are the 97.5th 
        and 2.5th percentile of probability distribution of 
        the estimated stellar mass; see Kauffmann et al. 
        2003b for more details). 
\end{enumerate}
The first criterion ensures the accurate measurement of 
$F$([O{\sc ii}]$\lambda$3727). The second criterion selects
emission-line galaxies. The S/N values are taken from the
emission-line flux catalog; however, the errors in this 
catalog are underestimated with a factor of $\sim$2 (see 
the web site for more details). 
The third criterion is required to 
reject AGNs from the sample, following Kauffmann et al. 
(2003a), although the AGN removal had been already examined 
in the process of checking the classification in the 
oxygen-abundance catalog mentioned above. The fourth 
criterion is required to avoid significant aperture effects 
on the flux ratios. Kewley et al. (2005) reported that the 
introduced systematic error of the metallicity determination 
reaches up to $\sim$40\% when the fiber covers only 20\% of 
the total $g^\prime$-band photons (see also Tremonti et al. 
2004). The fifth criterion is not directly relevant to this 
study, but is required in our companion paper (Nagao et al. 
2006b).

Note that in sample C we do not put constraints on
$\Delta$(log$\frac{\rm O}{\rm H})$. This is to prevent 
sample C to be devoid of galaxies at metallicities 
12+log(O/H) $<$ 7.6 so that sample C and sample A+B have 
some overlap in terms of metallicities. Indeed, due to the 
luminosity-metallicity relation, low metallicity galaxies 
in sample C are on average fainter and therefore tend to 
have larger errors. However, the averaged gas metallicities 
in sample C are most likely reliable even at low 
metallicities, thanks to the large number of objects in the 
sample.
Our final sample of emission-line galaxies consists of 48497
objects, which is hereafter called ``sample C''. The redshift 
distribution of galaxies in sample C is 
shown in Figure 4. Its median value is 0.085, while the mean 
and the RMS are 0.092 and 0.040, respectively.
The distribution of the oxygen abundance of galaxies in sample C
is shown in Figure 5. Its median value is 9.016, while the 
mean and the RMS are 8.976 and 0.166.
The means and the RMSs of the oxygen abundance of samples
A, B, and C are summarized in Table 2.

\section{Results}

In Figures 6 and 7, we plot emission-line flux ratios for 
the galaxies in samples A, B, and C. To avoid noisy objects 
in sample C, we consider only those 
with S/N $\geq$ 10 (cataloged value) for all the related 
emission lines (e.g., H$\beta$, [O{\sc ii}]$\lambda$3726, 
[O{\sc ii}]$\lambda$3729 and [O{\sc iii}]$\lambda$5007 for 
the case of $R_{23}$).
In addition to $R_{23}$, all the other flux ratios 
investigated here are metallicity-sensitive flux ratios and 
sometimes regarded as metallicity diagnostics (see, e.g., 
Kewley \& Dopita 2002; Pettini \& Pagel 2004; Kobulnicky \& 
Kewley 2004). Among them, 
$F$([O{\sc iii}]$\lambda$5007)/$F$([O{\sc ii}]$\lambda$3727)
is sensitive also to the ionization parameter and thus it
has {\it not} been regarded as a good metallicity diagnostic 
flux ratio (see Kewley \& Dopita 2002). Instead, this flux 
ratio has been used to investigate the ionization parameter, 
and has been sometimes used in the following form:
\begin{equation}  % equation (2)
  O_{32} = 
         \frac{ F({\rm [OIII]}\lambda4959) \! + \!
                F({\rm [OIII]}\lambda5007) } 
              { F({\rm [OII]}\lambda3727) }
\end{equation}
(e.g., Kobulnicky \& Kewley 2004). 

The data sequences in the diagnostics-metallicity diagrams
are mostly continuous for different samples, and
accordingly the whole sample shows clear relations between
various metallicity diagnostics and the oxygen abundance.
Since there are no apparent systematic differences in the
diagnostics-metallicity sequences between sample A and
sample B, we combined these two samples and identified as 
``sample A+B'' hereafter in order to improve the 
statistics at low metallicities.
The statistical properties of the sample A+B are given
in Table 2. The diagram of
$F$([N{\sc ii}]$\lambda$6584)/$F$([S{\sc ii}]$\lambda$6720)
versus the oxygen abundance shows an apparent 
discontinuity between sample A+B and sample C, where
$F$([S{\sc ii}]$\lambda$6720) denotes the sum of 
$F$([S{\sc ii}]$\lambda$6717) and 
$F$([S{\sc ii}]$\lambda$6731). We will discuss the issue 
of this discontinuity in \S\S4.1.

The diagram of $R_{23}$ versus the oxygen abundance shows
a $\cap$-shaped distribution with a peak at 12+log(O/H) 
$\sim$ 8.0. This is consistent with the previous studies 
on the {\it empirical} relation between $R_{23}$ and the 
oxygen abundance based on smaller samples of observational 
data (e.g., Edmunds \& Pagel 1984; McGaugh 1991; 
Miller \& Hodge 1996; Castellanos et al. 2002; 
Lee et al. 2003a; Bresolin et al. 2004, 2005; 
Pilyugin \& Thuan 2005). As discussed in \S\S4.3, however, 
this appears to be systematically different from
previous predictions of photoionization models.

To investigate the relation between the flux ratios and the
oxygen abundances quantitatively, we calculate the means 
and the RMSs of the flux ratios of galaxies within bins of 
oxygen abundance. For sample A+B, we calculate them in the 
range $7.05 <$ 12+log(O/H) $< 8.55$ with a bin width of 
$\Delta$[log(O/H)] = 0.1, except at lowest and highest 
oxygen abundances, where the bin width is wider (i.e., 
$7.05 <$ 12+log(O/H) $< 7.45$ and 
$8.35 <$ 12+log(O/H) $< 8.55$) due to the small number of 
sources in these ranges. All of the metallicity bins contain 
at least 6 galaxies. The results are given in Table 3. 
We also calculate the mean and the RMS of flux ratios for 
sample C in the range $8.15 <$ 12+log(O/H) $< 9.25$ with a 
bin width of $\Delta$[log(O/H)] = 0.1 dex. 
The results are given in Table 4.

The calculated mean and the RMS of the flux ratios for each 
metallicity bin are shown in Figures 8 and 9. We then fit 
the observed sequences between flux ratios and oxygen 
abundance with polynomial functions in the range 
$7.05 <$ 12+log(O/H) $< 9.25$ (or 
$0.02 < Z_{\rm gas}/Z_\odot < 4$), and the results of the 
fits are also shown in Figures 8 and 9. We decided to fit 
3rd-order polynomial functions for the binned data, not for 
the individual data, in order to avoid giving too much 
weights to the high metallicity range (where most of the 
data are). In Table 5, the coefficients of the 
best-fit polynomial functions are provided, 
according to the formula
\begin{equation}  % equation (3)
  {\rm log} \ (F_1 / F_2) =
  \sum_{N}^{}{a_N [{\rm log}(Z/Z_\odot)]^N}
\end{equation}
where $F_1 / F_2$ is the line flux ratio (or the $R_{23}$
parameter) and $N=$ (0, 1, 2, 3). 
% [we will discuss on 
% $F$([Ne{\sc iii}]$\lambda$3869)/$F$([O{\sc ii}]$\lambda$3727)
% in \S\S4.4]. 
Here we adopt 12+log(O/H)$_\odot$ = 8.69 for 
$Z_\odot$ (Allende Prieto et al. 2001).
For the convenience of the reader, in Table 6 we also give 
the coefficients of the best-fit polynomial functions
in the following form:
\begin{equation}  % equation (4)
  {\rm log} \ (F_1 / F_2) =
  \sum_{N}^{}{b_N [12 + {\rm log (O/H)}]^N}.
\end{equation}

The expected uncertainties on the derived metallicities from
the diagnostic flux ratios calibrated here can be estimated 
using the RMS values of the flux ratios given
for each mass bin (Tables 3 and 4).
By looking at the RMS plotted in Figures 8 and 9, 
we can recognize that, for instance, the diagnostic flux 
ratios give highly uncertain metallicities when
$F$([N{\sc ii}]$\lambda$6584)/$F$([O{\sc ii}]$\lambda$3727)
$\la$ 0.05,
$F$([O{\sc iii}]$\lambda$5007)/$F$([O{\sc ii}]$\lambda$3727)
$\ga$ 2, and
$F$([N{\sc ii}]$\lambda$6584)/$F$([S{\sc ii}]$\lambda$6720)
$\la$ 0.3. On the contrary, the metallicity is
well determined ($\Delta Z \la 0.2$ dex) when
$F$([O{\sc iii}]$\lambda$5007)/$F$([N{\sc ii}]$\lambda$6584)
$\la$ 10 and
$F$([N{\sc ii}]$\lambda$6584)/$F$([O{\sc ii}]$\lambda$3727)
$\ga$ 0.05.

Finally we recall that these relations are valid only 
in the range 7.05 $<$ 12+log(O/H) $<$ 9.25
(which is however much wider than in any previous work).
We warn on the use of these relations outside such
metallicity range, since it would rely only on their 
extrapolation.

\section{Discussion}

\subsection{Consistency between sample A+B and sample C}

Before interpreting the results, we discuss on the 
consistency of the two main samples, i.e., galaxies with 
(A+B) and without (C) [O{\sc iii}]$\lambda$4363 measurements. 
As mentioned in \S3, the relation between some emission-line 
flux ratios and the oxygen abundance is not smoothly 
connected between the two samples (A+B and C), and this is 
especially significant for the flux ratios of
$F$([O{\sc iii}]$\lambda$5007)/$F$([O{\sc ii}]$\lambda$3727)
and
$F$([N{\sc ii}]$\lambda$6584)/$F$([S{\sc ii}]$\lambda$6720),
but is also seen in other cases (Figures 8 and 9). One of 
the possible reasons for this discrepancy is a systematic 
error in the estimate of the oxygen abundance for one (or 
both) of the two different methods, which in one case 
consists in using the gas temperature inferred through 
[O{\sc iii}]$\lambda$4363 emission (see \S\S2.1) and in the 
other case is using all of optical strong emission lines 
(Tremonti et al. 2004). 

Kobulnicky et al. (1999) investigated a possible systematic 
error in the former method, that is, the gas temperature may 
be overestimated through the [O{\sc iii}]$\lambda$4363 
emission and thus the oxygen abundance may tend to be 
underestimated accordingly. This is because the strength of 
the [O{\sc iii}]$\lambda$4363 emission significantly depends 
on the gas temperature and thus spectra obtained by a 
global aperture toward a galaxy are biased towards higher 
gas-temperature H{\sc ii} regions (see also Peimbert 1967).
According to their analysis, the overestimation of the gas 
temperature could be more serious in low-metallicity 
systems and could reach up to 
$\Delta T_{\rm e} = 1000 - 3000$K, which results in the 
systematic underestimation of the oxygen abundance of 
0.05 -- 0.2 dex. However, although this effect may partly 
account for the discrepancy of the metallicity dependence of
$F$([N{\sc ii}]$\lambda$6584)/$F$([S{\sc ii}]$\lambda$6720),
it goes in the opposite direction to account for the 
discrepancy seen in 
$F$([O{\sc iii}]$\lambda$5007)/$F$([N{\sc ii}]$\lambda$6584)
and 
$F$([O{\sc iii}]$\lambda$5007)/$F$([O{\sc ii}]$\lambda$3727).
Therefore, the effect of the biased temperature measurement 
is not the dominant origin of the discontinuities seen in 
Figures 8 and 9.

A systematic error in the oxygen abundance may exist in the 
method of Tremonti et al. (2004). They estimated the oxygen 
abundance by comparing photoionization models with some 
optical emission-line fluxes, which were measured on the 
spectra after subtraction of the stellar component. Although 
their method of the stellar-component subtraction is a 
sophisticated one and uses the most recent population 
synthesis models of Bruzual \& Charlot (2003), it is not clear
whether the measurement of emission lines lying on the deep 
and complex stellar absorption features is completely free 
from some possible systematic errors. 
A possible improper subtraction of the stellar absorption 
features may lead to systematic errors on the fluxes of Balmer 
lines, which might result in a systematic error in the 
estimation of the gas metallicity of galaxies in sample C. 
The subtraction of stellar absorption features may be 
inaccurate also in sample A+B. For instance, in some earlier 
works the stellar subtraction was performed by simply assuming 
$EW$(H$\beta$)$_{\rm abs}$ = 1${\rm \AA}$. 
This over-simplified assumption may introduce systematic 
errors in the derived gas metallicity and the emission-line 
flux ratios given in Table 1.
Another possible source of uncertainty in the method of
Tremonti et al. (2004) is the use of the 
[N{\sc ii}]$\lambda$6584 flux and its comparison to models.
Most photoionization models assume that the relative 
nitrogen abundance scales with the metallicity linearly 
when the primary nitrogen creation dominates, and scales
quadratically when the secondary nitrogen creation is 
dominant. However, the transition metallicity between the 
two modes is uncertain. An inaccurate value of the 
transition metallicity (which is indeed uncertain) may 
lead to systematic errors in the
estimation of metallicity especially at low metallicities, 
which could be one of the possible origin of the discrepancy 
seen in Figures 8 and 9.

The discrepancy in the metallicity dependences of
emission-line flux ratios may also be a consequence of the
selection of spectroscopic targets. While galaxies in
sample C are basically selected in terms of their apparent 
magnitude and thus not largely biased toward any specific 
population, galaxies in sample B could be biased toward 
very strong emission-line galaxies (galaxies in sample A are
in a composite situation; see Izotov et al. 2006). 
This is because the 
motivation behind most of the original observations, such as 
the studies on the primordial helium abundance (see the 
original references given in Table 1), required very 
accurate measurements of emission-line flux ratios. For a 
given metallicity, galaxies with stronger emission lines 
tend to be characterized by a higher ionization parameter, 
which may result into larger flux ratios of
$F$([O{\sc iii}]$\lambda$5007)/$F$([N{\sc ii}]$\lambda$6584)
and
$F$([O{\sc iii}]$\lambda$5007)/$F$([O{\sc ii}]$\lambda$3727),
although the difference in the ionization parameter
should not cause a significant difference in the ratio of
$F$([N{\sc ii}]$\lambda$6584)/$F$([S{\sc ii}]$\lambda$6720).
We will discuss the effect of the ionization parameter
on the discrepancy further in \S\S4.3.

Actually some or all of the above matters could contribute 
to the discontinuity in the metallicity dependences of
emission-line flux ratios, and their discrimination or 
their accurate correction are not feasible. We thus simply 
adopt the results of the fit described in \S3 and not take 
the effects of the possible systematic errors into account 
in the following discussion. However, it should be noted 
that this rather complex situation is caused by relying
on two independent methods to measure the oxygen abundance.
This problem will be solved if a large sample of galaxies 
with a wide range of the oxygen abundance is investigated 
by using a unique method throughout the concerned 
metallicity range.

\subsection{Comparison with previous empirical calibrations}

We compare the results of our calibrations with previous 
empirical calibrations. In particular, in Figure 10, we
compare the empirical calibrations of $R_{23}$ derived
by us with those obtained by Tremonti et al. (2004), 
Edmunds \& Pagel (1984), and
Zaritsky et al. (1994). While there is a reasonable agreement
between our result and the result from previous calibration
for the lower branch (Edmunds \& Pagel 1984), there are
some systematic discrepancies for the upper branch.
We should in particular discuss the difference between our
calibration and that of Tremonti et al. (2004), since
our calibration in the high-metallicity range is based on
the metallicity of the SDSS galaxies (in sample C) derived by
Tremonti et al. (2004). 
The calibration by Tremonti et al. (2004), which is provided
only for the upper branch, is clearly flatter than ours.
This discrepancy may be ascribed to the combination of
various possible factors. Our calibration also includes the fit
of the new sample of [O{\sc iii}]$\lambda$4363-detected
galaxies, which are not included in Tremonti et al. (2004),
and this is certainly one of the reasons for the
discrepancy. However, the latter issue cannot completely
account for the discrepancy, since the Tremonti et al. (2004)
calibration fails to reproduce the SDSS data at 
12+log(O/H) $<$ 8.5 (as shown in Figure 10). It is likely
that an additional source of the discrepancy is the different
strategy of fitting the analytical function to the data.
While we fit the third polynomial function to the binned
data, Tremonti et al. (2004) fit the function to the whole
sample of individual SDSS galaxies. Since the number of high
metallicity galaxies [12+log(O/H) $>$ 8.5] is much larger
than the low metallicity sub-sample [12+log(O/H) $<$ 8.5]
as shown in Figure 5, the analytical fit of Tremonti 
et al. (2004) is dominated by the
high-metallicity part of the $R_{23}$ diagram. 
Finally, the modest discrepancy at high metallicities
[12+log(O/H) $>$ 9] may be partly attributed to the 
difference in the sample selection criteria. As described 
in \S3, we select the galaxies in the sample C with S/N $\geq$ 
10 for all of the lines H$\beta$, [O{\sc ii}]$\lambda$3726,
[O{\sc ii}]$\lambda$3729, and [O{\sc iii}]$\lambda$5007
(note that the [O{\sc ii}] doublet lines are measured 
separately in the original catalog we used), while
Tremonti et al. (2004) adopted the S/N criteria only 
for H$\beta$, H$\alpha$ and [N{\sc ii}]$\lambda$6584,
not for [O{\sc ii}]$\lambda$3726,
[O{\sc ii}]$\lambda$3729, and [O{\sc iii}]$\lambda$5007.
Our selection criteria may preferentially reject objects 
at high metallicities with respect to
those of Tremonti et al. (2004), because 
forbidden lines such as [O{\sc ii}] and [O{\sc iii}]
become weak when gas metallicity is high
due to the suppressed collisional excitation mechanism
(e.g., Ferland et al. 1984; Nagao et al. 2006a).
This effect may result in our selective choice of objects with 
strong [O{\sc ii}] and [O{\sc iii}] emission in a given
metallicity bin, which could make our calibration to be
steeper at high metallicities. Since the difference in
the calibration between ours and that of Tremonti et al. (2004)
is significant at 12+log(O/H) $>$ 9, it is suggested that
our calibration for the $R_{23}$ may overestimate the gas 
metallicity at 12+log(O/H) $>$ 9 by a factor of 
$\Delta Z \sim 0.1$ dex at 12+log(O/H) $\sim$ 9.1.

The calibration of the diagnostic flux ratio
$F$([N{\sc ii}]$\lambda$6584)/$F$(H$\alpha$) is especially
important, because wavelength separation of the two lines
is small (i.e., not sensitive to dust reddening and 
requiring only small wavelength coverage) and thus
it is used as a diagnostic of the gas metallicity of galaxies
at $z \la 2$ (e.g., Erb et al. 2006).
In Figure 11, we compare the empirical calibrations of
$F$([N{\sc ii}]$\lambda$6584)/$F$(H$\alpha$) derived
by us, with those derived by Pettini \& Pagel (2004) and 
Denicol\'{o} et al. (2002). Our result agree reasonably well 
with Denicol\'{o} et al. (2002) only at sub solar metallicity, 
and there is a systematic difference in the slope between our 
result and the result reported by Pettini \& Pagel (2004). The
latter discrepancy may be due to the reduced
metallicity range of the sample of Pettini \& Pagel (2004), 
indeed most of their objects
are distributed within 7.7 $<$ 12+log(O/H) $<$ 8.5.
However, the difference is significant ($\Delta Z \ga 0.2$ dex)
only at metallicities 12+log(O/H) $<$ 7.5 and 
12+log(O/H) $>$ 8.5. Although the difference in the 
lowest-metallicity range is not a serious problem (because
in this metallicity range the expected [N{\sc ii}]$\lambda$6584
flux is extremely weak and thus its measurement would be
very challenging and probably inaccurate), 
it is important to pay attention to the difference in 
the high-metallicity range. Note that such ``high-metallicity'' 
range (where the discrepancy with Pettini \& Pagel 2004 occurs) 
is not so metal rich --- the metallicity 12+log(O/H) = 8.5
corresponds to $Z = 0.65 Z_\odot$, still in the sub-solar 
metallicity domain.

In Figure 12, we compare the empirical calibrations of
$F$([O{\sc iii}]$\lambda$5007)/$F$([N{\sc ii}]$\lambda$6584
derived by us with the one derived by Pettini \& Pagel (2004).
The difference between the two calibrations is more serious
than that seen in Figure 11 in the low metallicity range,
12+log(O/H) $<$ 8. However Pettini \& Pagel (2004) 
correctly mentioned that the flux ratio 
$F$([O{\sc iii}]$\lambda$5007)/$F$([N{\sc ii}]$\lambda$6584
is of little use when 
$F$([O{\sc iii}]$\lambda$5007)/$F$([N{\sc ii}]$\lambda$6584
$\ga$ 100 because of the saturation of this diagnostic. 
The behavior of this diagnostic flux ratio in
the low-metallicity range would be important to
derive the upper limits on the metallicity from an
upper limit of the [N{\sc ii}]$\lambda$6584 flux.

\subsection{Comparison with photoionization models}

To interpret the metallicity dependences of the 
emission-line flux ratios, we compare observational data
with the predictions of 
photoionization models. In Figures 13 and 14, we show 
the {\it empirical} metallicity dependences and the 
{\it theoretical} metallicity dependences of some 
metallicity diagnostics, where the latter are taken from 
Kewley \& Dopita (2002) except for 
$F$([N{\sc ii}]$\lambda$6584)/$F$(H$\alpha$) that is 
taken from Kobulnicky \& Kewley (2004). Since the explicit 
analytic expression for the metallicity dependence of 
$F$([O{\sc iii}]$\lambda$5007)/$F$([O{\sc ii}]$\lambda$3727)
is not given by Kewley \& Dopita (2002), we derive the 
polynomial expression of the theoretical metallicity 
dependence by fitting the results given in Table 2 of
Kewley \& Dopita (2002). The photoionization models 
presented by Kewley \& Dopita (2002) and Kobulnicky \& 
Kewley (2004) were calculated by the photoionization code 
MAPPINGS III (Sutherland \& Dopita 1993) combined with the 
stellar population synthesis codes PEGASE (Fioc \& 
Rocca-Volmerange 1997) and STARBURST99 (Leitherer et al. 
1999), for the range $7.6 <$ 12+log(O/H) $< 9.4$. 
They assume that stars and gas have the same metallicity,
which is a reasonable assumption given that 
photoionization is due to hot, young stars, presumably 
recently formed from the same gas that they are 
photoionizing. In their 
calculations, nitrogen is assumed to be a secondary 
nucleosynthesis element at 12+log(O/H) $>$ 8.3, and a 
primary nucleosynthesis element at lower metallicity. 
Effects of dust grains on the depletion of gas-phase heavy 
elements and on the radiative transfer are consistently 
taken into account. Their calculations cover the range of 
ionization parameters $-3.8 \leq {\rm log} \ U \leq -2.0$, 
or equivalently, $5\times10^6 {\rm cm \ s}^{-1} \leq q 
\leq 3\times10^8 {\rm cm \ s}^{-1}$ (where $U \equiv q/c$). 
See Kewley \& Dopita (2002) for details on the calculations. 
Note that they adopted 12+log(O/H)$_\odot$ = 8.93 
(Anders \& Grevesse 1989) and expressed the metallicity in 
units of $Z_\odot$ (12+log(O/H)$_\odot$ = 8.93). However, 
since we adopt a more recent value for the solar abundance, 
12+log(O/H)$_\odot$ = 8.69 (Allende Prieto et al. 2001),
the metallicity notation is different when the 
$Z_\odot$ unit is used, which should be kept in mind to 
compare our results with their predictions.

The most remarkable matter in the comparison between the 
empirical and theoretical metallicity dependences of
emission-line flux ratios is the significant discrepancy
in the theoretically-expected $R_{23}$-sequence with respect 
to the observed trend. This is especially significant at 
low metallicity range 12+log(O/H) $<$ 8. Shi et al. (2006)
also recently reported that a previous theoretical calibration
of $R_{23}$ (see McGaugh 1991; Kobulnicky et al. 1999)
overpredicts the gas metallicity with respect to the
metallicity measured through the gas temperature
determined with [O{\sc iii}]$\lambda$4363 line 
($\Delta Z \sim 0.2$ dex), especially in the low metallicity 
range [i.e., 12+log(O/H) $<$ 8]. This discrepancy is not due 
to an improper compilation in our data, because it has 
been reported also in the earlier works that the 
empirical peak of $R_{23}$ is seen around 12+(O/H) $\sim$ 
8.0, as mentioned already in \S3. The discrepancy cannot 
be ascribed to problems to the model results of 
Kewley \& Dopita (2002) either, because other theoretical works 
also predict higher peak metallicity of $R_{23}$ 
independently [12+log(O/H) $\ga$ 8.3; e.g.,  
Kobulnicky et al. 1999]. One possible idea to reconcile 
this discrepancy is that the ionization parameter of the gas 
is higher than the parameter range which Kewley \& Dopita 
(2002) covers, especially in low-metallicity objects. If 
the ionization parameter correlates negatively with the 
gas metallicity and it reaches up to log $U > -2$ at the 
lowest metallicities, photoionization models would predict 
larger values of $R_{23}$ in the lower-metallicity range
with respect to constant-$U$ models. This 
idea appears to be consistent with the behaviors of the 
empirical sequences in the $U$-sensitive flux ratios,
$F$([O{\sc iii}]$\lambda$5007)/$F$([N{\sc ii}]$\lambda$6584)
(Figure 13) and 
$F$([O{\sc iii}]$\lambda$5007)/$F$([O{\sc ii}]$\lambda$3727)
(Figure 14). By focusing on these two $U$-sensitive flux 
ratios, we can see that the ionization parameter increases 
by $\sim$0.7 dex with decreasing oxygen abundance from 
12+log(O/H) = 9.0 to 7.5, supporting the above 
interpretation. Although the absolute value of the 
required ionization parameter appear to be inconsistent 
between $R_{23}$ and the latter two $U$-sensitive flux 
ratios, the inferred absolute $U$ values depends also on 
some model assumptions such as the spectral energy distribution
(SED) of ionizing photons 
or the relative elemental abundance ratios, which also 
change as a function of metallicity. We thus 
conclude that the metallicity dependence of the ionization 
parameter (hereafter ``$U$-$Z$ relation'') causes the 
discrepancy between the empirical 
$R_{23}$ distribution and the model predictions with a 
constant ionization parameter.

Note that the
$F$([O{\sc iii}]$\lambda$5007)/$F$([N{\sc ii}]$\lambda$6584)
and 
$F$([O{\sc iii}]$\lambda$5007)/$F$([O{\sc ii}]$\lambda$3727)
ratios are also sensitive to the hardness of the ionizing
radiation, which is a strong function of the stellar
metallicity. This effect can in principle also contribute
to the dependence of 
$F$([O{\sc iii}]$\lambda$5007)/$F$([N{\sc ii}]$\lambda$6584)
and 
$F$([O{\sc iii}]$\lambda$5007)/$F$([O{\sc ii}]$\lambda$3727)
ratios on metallicity. However, the models by Kewley \& 
Dopita (2002) plotted in Figures 13 and 14 already take into 
account the hardening of the stellar spectra as a function
of metallicity. Therefore, the discrepancy between 
constant-$U$ models and the data indicates that the 
hardening of the ionizing spectra must be associated with a
variation of $U$ with metallicity. In particular, the
dependence of the 
$F$([O{\sc iii}]$\lambda$5007)/$F$([O{\sc ii}]$\lambda$3727)
ratio on metallicity cannot entirely be ascribed only
to the hardening of the ionizing radiation, but also to a
$U$-$Z$ relation.

The inferred $U$-$Z$ relation is a very interesting 
result. Maier et al. (2006) also recently reported that 
the lower-metallicity galaxies tend to be characterized
by a higher ionization parameter (see also 
Maier et al. 2004, who reported the correlation between the 
absolute $B$ magnitude and the flux ratio of
$F$([O{\sc iii}]$\lambda$5007)/$F$([O{\sc ii}]$\lambda$3727)
among galaxies in the local universe).
Although a detailed theoretical interpretation of this empirical 
relation goes beyond the scope of this paper, in the 
following we discuss two possible qualitative interpretations. 
One possible origin of this effect may be associated with the 
mass--metallicity relation and with the mass--age relation in 
local galaxies. According to these relations, higher 
metallicity galaxies are associated with more massive and 
older systems. H{\sc ii} regions ionized by later stellar 
populations are expected to be characterized by lower 
ionization parameters, due to the lower luminosity of the 
ionizing stars. Another possible explanation may be the
(plausible) relation between gas metallicity and stellar 
metallicity, and in particular that lower metallicity gas is 
ionized by lower metallicity stars. For a given stellar mass, 
lower metallicity stars emit a harder and stronger radiation
field, therefore giving a higher ionization parameter. The 
latter effect would naturally yield a $U$-$Z$ relationship. 
The former are just qualitative interpretations. However, a 
thorough investigation of this phenomenon will requite detailed 
observational studies of stellar population in star forming 
galaxies.

The comparison of the empirical and the theoretical 
sequences of the two $U$-sensitive diagnostic flux ratios,
$F$([O{\sc iii}]$\lambda$5007)/$F$([N{\sc ii}]$\lambda$6584) 
and 
$F$([O{\sc iii}]$\lambda$5007)/$F$([O{\sc ii}]$\lambda$3727),
also suggests the fact that the dispersion of the
ionization parameter for a given metallicity should be
relatively small. The typical RMS of the two flux ratios 
are $\sim$0.5 (in logarithm) at 12+log(O/H) $\sim$ 7.5.
This corresponds to an RMS of the ionization parameter of
$\sim$0.5 dex. This is the reason why the very 
$U$-sensitive flux ratio,
$F$([O{\sc iii}]$\lambda$5007)/$F$([O{\sc ii}]$\lambda$3727),
shows a clear metallicity dependence as seen in Figure 13.
The $U$-metallicity relationship is also important to
understand the behavior of the empirical metallicity 
dependence of the flux ratio
$F$([O{\sc iii}]$\lambda$5007)/$F$([N{\sc ii}]$\lambda$6584).
This flux ratio is predicted to decrease with
the oxygen abundance below 12+log(O/H) $\sim$ 7.6 by 
photoionization models with a constant ionization 
parameter. Owing to the metallicity dependence 
of the ionization parameter, this flux ratio does not 
show the ``turnover'' seen in $R_{23}$ and thus it is very 
useful to investigate the gas metallicity of galaxies 
without the measurement of $F$([O{\sc iii}]$\lambda$4363). 
Another implication of these results is that one should 
not use constant-$U$ photoionization models to derive the oxygen 
abundance from the observed flux ratios, not only from
$F$([O{\sc iii}]$\lambda$5007)/$F$([N{\sc ii}]$\lambda$6584
but also from any other metallicity diagnostics, which 
introduce systematic errors in the calibration. 
The empirical relations provided in 
this paper (Tables 5 and 6) are very useful to avoid such
systematic errors to derive the gas metallicity by using 
only strong emission lines.

As for the $U$-insensitive diagnostic flux ratios,
$F$([N{\sc ii}]$\lambda$6584)/$F$(H$\alpha$),
$F$([N{\sc ii}]$\lambda$6584)/$F$([O{\sc ii}]$\lambda$3727) 
and
$F$([N{\sc ii}]$\lambda$6584)/$F$([S{\sc ii}]$\lambda$6720),
there are no significant discrepancies between the 
empirical sequence and the theoretical sequence (with a 
constant ionization parameter). This indirectly supports 
the above interpretation that the apparent discrepancy 
in $R_{23}$ between the empirical sequence and the 
results of photoionization model is caused by the effect 
of the ionization parameter. Note that there is little 
or no metallicity dependence of the flux ratios of
$F$([N{\sc ii}]$\lambda$6584)/$F$([O{\sc ii}]$\lambda$3727) 
and
$F$([N{\sc ii}]$\lambda$6584)/$F$([S{\sc ii}]$\lambda$6720)
in the low-metallicity range, 12+log(O/H) $\la$ 8.0, in terms 
both of empirical and theoretical dependences. Therefore
these diagnostic flux ratios are useful only for the 
high metallicity galaxies.

The photoionization models presented in Figures 13 and 14
suggest an additional interpretation of the discrepancy 
in some diagnostics between the two samples discussed in 
\S\S4.1 (i.e., the discontinuity between sample A+B and 
sample C). Focusing on the metallicity range of 12+log(O/H) 
$\sim$ 8.3 where the two datasets of sample A+B and sample C 
overlap, we note that the trend of the discrepancy suggests 
that the galaxies in sample A+B have higher ionization 
parameter than the galaxies in sample C. This supports the 
interpretation that the discrepancy is at least partly 
caused by the selection effect, i.e., galaxies with higher 
ionization parameter are selectively picked up in sample A+B. 
Then, what causes this selection effect? This may be related 
with the fact that the [O{\sc iii}]$\lambda$4363 emission is
extremely weak in higher metallicity galaxies. This means 
that we can measure the [O{\sc iii}]$\lambda$4363 flux of 
galaxies with 12+log(O/H) $\sim$ 8.3 (the highest 
metallicity in the galaxies in sample A+B) only when
the [O{\sc iii}] emission is very strong, which corresponds
to a very high ionization parameter.

\subsection{Implications for studies of high-redshift galaxies
            and new diagnostics}

Although the $R_{23}$ method is thought to be a good
metallicity diagnostic, various other diagnostics (some 
of which are investigated in this paper) have been proposed 
up to now. Indeed one of the main problems of the $R_{23}$ 
method is that there are two solutions for a given $R_{23}$ 
value and thus one cannot obtain a unique metallicity solution. 
Most of the newly proposed diagnostics use the 
[N{\sc ii}]$\lambda$6584 line to remove the degeneracy 
in $R_{23}$, because the secondary nucleosynthesis of 
nitrogen makes this line emission very sensitive to the 
gas metallicity. However, there are two non-negligible
problems with the use of the [N{\sc ii}]$\lambda$6584 line.
First, especially for low-metallicity systems, the
contribution of the primary nucleosynthesis and the
secondary nucleosynthesis in the nitrogen abundance is
not well understood, which leads to an uncertainty in 
the relative nitrogen abundance as a function of the metallicity.
Second, the [N{\sc ii}]$\lambda$6584 emission is in the
red part of the rest-frame optical spectrum 
of galaxies, which prevents its application to the 
observational investigations of high-$z$ systems.
For example, the optical detectors with a sensitivity
up to $\lambda \sim 1\mu$m can detect the
[N{\sc ii}]$\lambda$6584 emission of galaxies only at 
$z \la 0.52$, and the $K$-band atmospheric window 
limits the highest redshift to $z \sim 2.7$ for 
ground-based facilities. Although one of the 
undoubtfully interesting targets for the JWST is the 
population related to the cosmic reionization, the 
sensitivity of NIRSpec (Posselt et al. 2004)
boarded on JWST can examine the 
[N{\sc ii}]$\lambda$6584 emission of the objects at 
$z \la 6.6$, where the cosmic reionization has 
already nearly ended (e.g., Kashikawa et al. 2006; 
Fan et al. 2006).
Another problem associated with the [N{\sc ii}]$\lambda$6584 
line is that it becomes very weak and difficult to measure at
low metallicities: [N{\sc ii}]$\lambda$6584/H$\alpha$ $<$ 0.1
at 12+log(O/H) $<$ 8.5.

Our results on the empirical metallicity dependences 
suggest that one does not need [N{\sc ii}]$\lambda$6584 
any more to distinguish the upper- and lower-branches 
of the $R_{23}$ sequence. This is because the flux 
ratio of 
$F$([O{\sc iii}]$\lambda$5007)/$F$([O{\sc ii}]$\lambda$3727)
is also a good metallicity diagnostics, thanks to the
small dispersion of the ionization parameter at a given
metallicity. The empirical $R_{23}$ sequence 
peaks at 12+log(O/H) $\sim$ 8.0, where the 
empirically determined flux ratio of 
$F$([O{\sc iii}]$\lambda$5007)/$F$([O{\sc ii}]$\lambda$3727)
is $\sim$2. Therefore one can recognize whether the observed 
$R_{23}$ belongs to the upper-branch of the $R_{23}$ 
sequence or not, depending on whether
$F$([O{\sc iii}]$\lambda$5007)/$F$([O{\sc ii}]$\lambda$3727)
$<$ 2 or not.
Note that this result is consistent with an earlier remark 
by Maier et al. (2004) that the flux ratio of
$F$([O{\sc iii}]$\lambda$5007)/$F$([O{\sc ii}]$\lambda$3727)
can be used to distinguish the upper- and lower-branches 
of the $R_{23}$ sequence. Our work gives the physical
explanation for this idea (the $U$-$Z$ relation) and
a criterion to distinguish the degeneracy
[$F$([O{\sc iii}]$\lambda$5007)/$F$([O{\sc ii}]$\lambda$3727)
$<$ 2] on the remark by Maier et al. (2004).

The above result is due to the fact that the ionization
parameter has a strong metallicity dependence, and it 
thus implies that the ionization parameter itself is a 
sort of metallicity diagnostic. Motivated by this, we
examine the metallicity dependence of the flux ratio
$F$([Ne{\sc iii}]$\lambda$3869)/$F$([O{\sc ii}]$\lambda$3727),
in Figure 15. The reasons for focusing on this flux ratio
are: (a) the two emission lines have different ionization
degrees, their ratio should have a strong dependence on the 
ionization parameter and therefore is a possible 
good metallicity diagnostics; (b) their wavelength 
separation is very small and thus their flux ratio is not 
significantly affected by dust reddening; and (c) the two 
lines are located at a blue part in the rest-frame optical 
spectrum and thus their flux ratio could be a powerful
diagnostic even for high-$z$ galaxies. As expected, this 
flux ratio shows a clear metallicity dependence, which is 
apparently seen in Figure 14. 
In Tables 7 and 8, the mean and the RMS of this
flux ratio for within each bins of oxygen abundance are
given, just similar to Tables 3 and 4 (\S3).
To obtain the analytic 
expression of this relation, we fit the observed sequence 
with a second-order polynomial function. 
The coefficients of the fit are given in Tables 9 and 10.

This flux ratio can be measured for galaxies up to 
$z \sim 1.6$ with optical instruments, up to $z \sim 5.2$ with 
near-infrared instruments on the ground-based facilities, and 
up to $z \sim 12$ with JWST/NIRSpec, 
therefore this flux ratio is a promising 
tool for metallicity studies at high redshift. 
In particular, it is useful for low metallicity 
galaxies, for which the intensity of [Ne{\sc iii}]$\lambda$3869 
becomes comparable to [O{\sc ii}]$\lambda$3727 and therefore 
easier to detect 
[$F$([Ne{\sc iii}]$\lambda$3869)/$F$([O{\sc ii}]$\lambda$3727) 
$>$ 0.2 at 12+logO/H $<$8]. 
Detailed theoretical calibrations 
on this flux ratio are required, taking the metallicity 
dependence of the ionization parameter into account,
which go beyond the scope of this paper.

One possible caveat for the use of the diagnostic flux 
ratios of
$F$([Ne{\sc iii}]$\lambda$3869)/$F$([O{\sc ii}]$\lambda$3727)
[and
$F$([O{\sc iii}]$\lambda$5007)/$F$([O{\sc ii}]$\lambda$3727),
too] may be the effect of AGNs. Since AGNs also tend to 
show higher ratios of 
$F$([Ne{\sc iii}]$\lambda$3869)/$F$([O{\sc ii}]$\lambda$3727)
and
$F$([O{\sc iii}]$\lambda$5007)/$F$([O{\sc ii}]$\lambda$3727),
galaxies harboring an AGN may be misidentified as 
low-metallicity galaxies. However, we can identify AGNs
through the detection of
He{\sc ii}$\lambda$4686 and/or [Ne{\sc v}]$\lambda$3426. 
Nagao et al. (2001) reported that typical type-2 AGNs show
$F$([Ne{\sc v}]$\lambda$3426)/$F$([O{\sc ii}]$\lambda$3727)
$\sim$0.4, and typical type-1 AGNs show even higher ratio
($\ga 1$). 
Lamareille et al. (2004) also reported that AGNs and
star-forming galaxies can be distinguished by using 
diagnostic diagrams using only the blue part of the
spectrum, i.e., $O_{32}$ versus $R_{23}$ and 
$F$([O{\sc iii}]$\lambda$5007)/$F$(H$\beta$) versus 
$F$([O{\sc ii}]$\lambda$3727)/$F$(H$\beta$) (see also
Rola et al. 1997). These suggest that we can easily distinguish 
AGNs from low-metallicity galaxies by using only
diagnostics available in the blue part of the spectrum,
even with moderate quality spectroscopic data.
Another caveat for the use of some diagnostic flux ratios
calibrated in this paper especially for high-$z$ galaxies
is that several of the empirical relations rely on the
$U$-$Z$ relation. It is not obvious that the $U$-$Z$ 
relation found in the local galaxies also holds for 
high-$z$ galaxies. If the $U$-$Z$ relation is a consequence 
of the relation between gas and stellar metallicity, as 
discussed in the previous section, then the relation is not 
expected to evolve and should remain valid at any redshift. 
Instead, if the $U$-$Z$ relation is a consequence of the 
mass-metallicity relation which evolves with redshift
(Savaglio et al. 2005; Erb et al. 2006; see also Maier et al. 
2004), then also the $U$-$Z$ relation may evolve with redshift 
and may require a re-calibration of 
our empirical relations at high redshift. The latter case 
would be a serious problem for several studies at high 
redshift. Indeed, most of the gas metallicity diagnostics 
discussed in this paper, including the ones most widely 
used (e.g. $R_{23}$), are significantly affected by the 
dependence on the ionization parameter.

Another difficulty to measure the gas metallicities of
high-$z$ galaxies is the faintness of targets, which
sometimes prevents from measuring accurate emission-line
fluxes. The use of low-resolution grating to improve the 
signal-to-noise ratio may yield to a blending of the
H$\alpha$ and [N{\sc ii}] emission lines, which results in 
poor determinations of the gas metallicity. Therefore It 
may be useful to investigate metallicity diagnostics which 
use the sum of $F$(H$\alpha$) and $F$([N{\sc ii}]). In
particular, we have examined the metallicity dependence of 
the flux ratio
$F$(H$\alpha$+[N{\sc ii}]$\lambda\lambda$6548,6584)/$F$([S{\sc ii}]$\lambda$6720)
in Figure 16. This group of lines can be measured even in 
low-resolution spectra and even in spectra covering a
relatively narrow wavelength range, and therefore may be
particularly useful in high-$z$ studies.There is a clear 
dependence of this flux ratio
on the oxygen abundance, seen as a $\cup$-shaped distribution
with a minimum at 12+log(O/H) $\sim$ 8.7 (i.e., $Z_{\rm gas} 
\sim Z_\odot$). The mean and the RMS of this flux ratio for 
each bin of oxygen abundance are given in Tables 7 and 8, and 
the coefficients of the fit are given in Tables 9 and 10.
The observed distribution of this flux ratio is naturally
expected, since the behavior of the nitrogen emission as a 
secondary element should dominate at the super-solar 
metallicity range, while $F$([N{\sc ii}]) and $F$([S{\sc ii}]) 
should become weak with respect to $F$(H$\alpha$) at the 
low-metallicities due to the decrease of the corresponding 
ions. Although this diagnostic like $R_{23}$ has two solutions 
when the ratio is below 10, this ratio seems useful for 
low-metallicity galaxies where it is larger than 10, in which 
case it is possible to state that the object belongs to the 
lower branch of the $\cup$-shaped distribution.
This diagnostic is also useful when the [S{\sc ii}] emission
is not detected (this is frequently the case when high-$z$
faint galaxies are concerned). In this case, we can calculate
a lower limit for this flux ratio, and we can derive 
accordingly an upper limit to the gas metallicity if the
lower limit is larger than 10. Note that this diagnostic is 
essentially independent of dust reddening.

Finally, We have also investigated the flux ratios of
$F$([O{\sc iii}]$\lambda$5007)/$F$(H$\beta$) and
$F$([O{\sc ii}]$\lambda$3727)/$F$(H$\beta$) (Figures 17 
and 18). The empirical calibrations for these two flux 
ratios may be useful when either [O{\sc ii}] or
[O{\sc iii}] are not available, because out of the 
wavelength range or on a strong OH airglow emission line, 
or in a region of bad atmospheric transmission.
The means and the RMSs of these two flux ratios for 
each bin of oxygen abundance are given in Tables 7 and 8, 
and the coefficients of the fit are given in Tables 9 
and 10. As expected, both of the two flux ratios again 
show $\cap$-shaped distributions just similar to the
$R_{23}$ parameter. 
The $F$([O{\sc iii}]$\lambda$5007)/$F$(H$\beta$) ratio 
is useful especially for high metallicity galaxies,
because the targets should belong to the upper-branch 
of the distribution when 
$F$([O{\sc iii}]$\lambda$5007)/$F$(H$\beta$) $<$ 1
(although this might be wrong when extremely
metal-poor galaxies [12+log(O/H) $<$ 7.0] are concerned).
Note that this flux ratio is very sensitive to the
oxygen abundance and the dispersion of the data is
small at $F$([O{\sc iii}]$\lambda$5007)/$F$(H$\beta$) 
$<$ 1.

Figure 19 summarizes the use of some of the metallicity
diagnostics discussed in this paper as a function of
redshift and for various observing facilities, and in
particular optical spectrometers, ground-based near-IR
spectrometers and NIRSpec on board of JWST. 
In principle (i.e., sensitivity permitting), MIRI
on board of JWST will be able to observe
the same diagnostics at even higher redshifts.
Note that the ratio
$F$([Ne{\sc iii}]$\lambda$3869)/$F$([O{\sc ii}]$\lambda$3727)
extends the diagnostic capability of any observing
facility to significantly higher redshift.

\section{Summary}

We have combined two large spectroscopic datasets to derive 
empirical calibrations for gas metallicity diagnostics 
involving strong emission lines. The two datasets consist 
of about 50000 spectra from the SDSS DR4, which probe 
metallicities 12+log(O/H)$>$8.3 (sample C), and of 328 
spectra of low metallicity galaxies with a measurement of the 
[O{\sc iii}]$\lambda$4363 line (sample A+B), which probe 
metallicities 12+log(O/H)$<$8.4. Together, these two 
samples provide the largest dataset of galaxies with known
metallicity currently available, and spanning more
than 2 dex in metallicity.

We have provided empirical calibrations both for 
metallicity diagnostics already proposed in the past and 
for new metallicity indicators proposed in this paper.
We have given an analytical description for 
the metallicity dependence of the following diagnostics 
and line ratios: $R_{23}$,
$F$([N{\sc ii}]$\lambda$6584)/$F$(H$\alpha$),
$F$([O{\sc iii}]$\lambda$5007)/$F$([N{\sc ii}]$\lambda$6584),
$F$([N{\sc ii}]$\lambda$6584)/$F$([O{\sc ii}]$\lambda$3727),
$F$([N{\sc ii}]$\lambda$6584)/ $F$([S{\sc ii}]$\lambda$6720),
$F$([O{\sc iii}]$\lambda$5007)/$F$([O{\sc ii}]$\lambda$3727), and
$F$([Ne{\sc iii}]$\lambda$3869)/$F$([O{\sc ii}]$\lambda$3727).
The calibrations are performed within the metallicity
range 7.0 $\leq$ 12+log(O/H) $\leq$ 9.2.
All of the investigated flux ratios show strong 
dependences on metallicity, 
at least in some metallicity ranges.
We have shown that the monotonic metallicity dependence 
of the ratio
$F$([O{\sc iii}]$\lambda$5007)/$F$([O{\sc ii}]$\lambda$3727)
can be used to break the degeneracy of the $R_{23}$ 
parameter when $F$([N{\sc ii}]$\lambda$6584)/$F$(H$\alpha$)
is not available. The
$F$([O{\sc iii}]$\lambda$5007)/$F$([O{\sc ii}]$\lambda$3727)
ratio is particularly useful at high redshift, where
H$\alpha$ and [N{\sc ii}]$\lambda$6584 are shifted outside 
the observed band. Another promising metallicity tracer at
high-$z$ is the ratio 
$F$([Ne{\sc iii}]$\lambda$3869)/$F$([O{\sc ii}]$\lambda$3727),
which is found to anti-correlate with metallicity. The 
$F$([Ne{\sc iii}]$\lambda$3869)/$F$([O{\sc ii}]$\lambda$3727),
ratio is particularly useful at high redshift, where most 
of the other diagnostic lines are shifted outside
the observed band.

We have also investigated the observed relationships 
through a comparison with photoionization models. Some of 
the diagnostics investigated in this paper are strongly 
dependent on the ionization parameter $U$. The observed 
trends of these diagnostics highlight a clear, inverse
relationship between ionization parameter and metallicity 
in galaxies. Such a strong $U$-$Z$ relationship is also 
required to explain the trend observed for the
$R_{23}$ parameter. The $U$-$Z$ relationship is relatively 
tight and, indeed, we have found that at any given metallicity 
the ionization parameter has a small dispersion 
($\sim$0.5~dex). The strong relationship between 
ionization parameter and metallicity in galaxies should 
warn about the use of simple models, which assume constant 
ionization parameter, to infer gas metallicities from line 
ratios.

\begin{acknowledgements}
  We thank J. Lee for comments on the flux data of the KISS 
  galaxies, Y. I. Izotov for comments on the relation between 
  $T_{\rm e}$(O$^+$) and $T_{\rm e}$(O$^{2+}$), L. Kewley
  for providing us their model results, and C. Maier,
  M. Onodera, and the anonymous referee for useful comments 
  on this work. 
  This work is based on the SDSS data catalogs released
  from Max Planck Institute for Astrophysics (MPA) and
  John Hopkins University (JHU), and produced by
  S. Charlot, G. Kauffmann, S. White, T. Heckman, 
  C. Tremonti, and J. Brinchmann.
  TN acknowledges financial 
  support from the Japan Society for the Promotion of Science 
  (JSPS) through JSPS Research Fellowship for Young Scientists. 
  RM and AM acknowledge financial support from the Italian
  Space Agency (ASI) and the Italian Institute for Astrophysics
  (INAF).
\end{acknowledgements}

\clearpage

\begin{figure}
\centering
\rotatebox{-90}{\includegraphics[width=8.0cm]{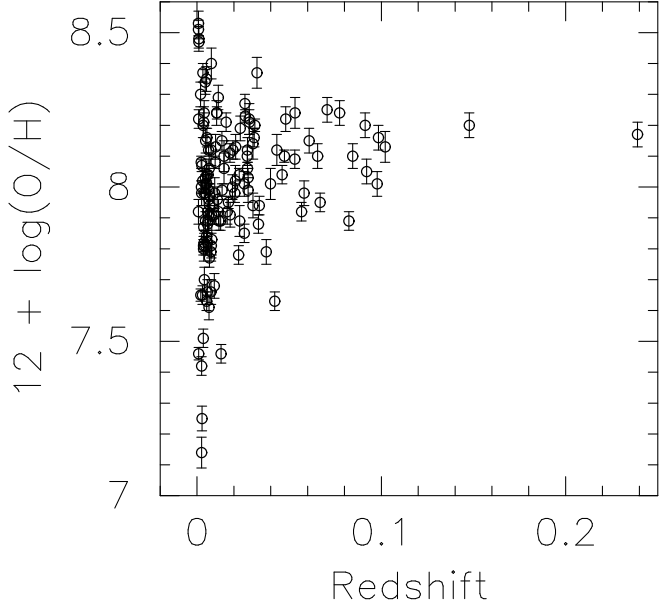}}
\caption{
   Oxygen abundance of the galaxies in sample A,
   derived by Izotov et al. (2006b), as a function of redshift.
}
\label{fig01}
\end{figure}

\begin{figure}
\centering
\rotatebox{-90}{\includegraphics[width=8.0cm]{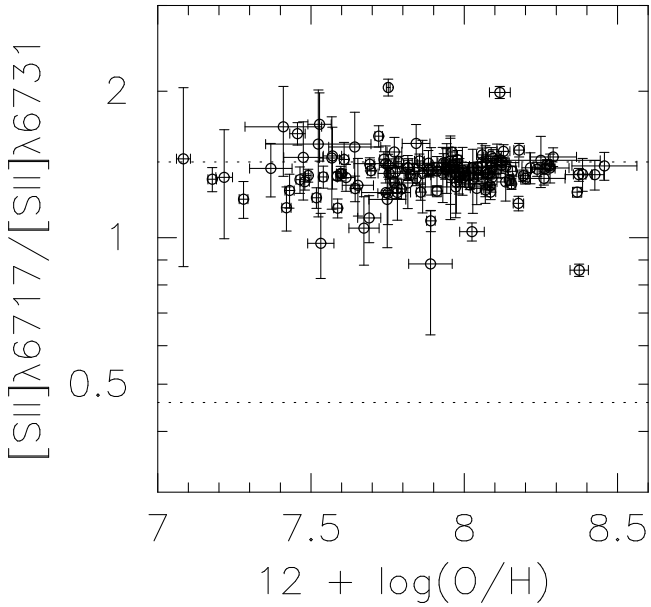}}
\caption{
   Emission-line flux ratios of 
   [S{\sc ii}]$\lambda$6717/[S{\sc ii}]$\lambda$6731 for
   the compiled low-metallicity galaxies, as a function of
   the oxygen abundance derived by us (see \S\S2.1.2). 
   The upper horizontal
   dotted line denotes the theoretical low-density limit
   of this flux ratio and the lower dotted line denotes the
   high-density limit. 
}
\label{fig02}
\end{figure}

\begin{figure}
\centering
\rotatebox{-90}{\includegraphics[width=6.8cm]{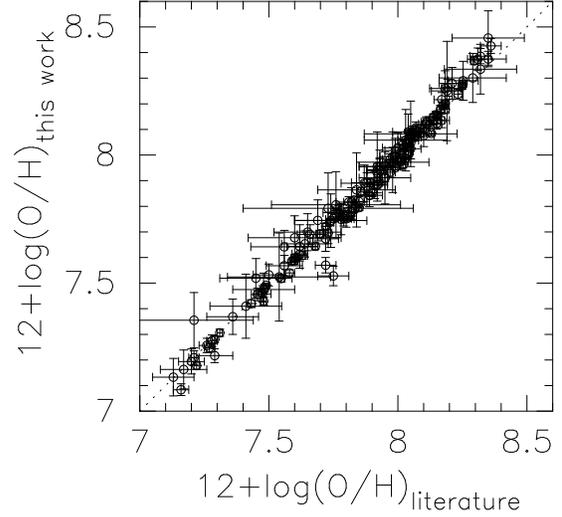}}
\caption{
   Oxygen abundances of the compiled low-metallicity galaxies
   re-calculated by us (see \S\S2.1.2) are plotted as a
   function of the oxygen abundances given in the original
   references. Dotted line is not the best-fit line but
   a reference line for the case when the two quantities
   are the same.
}
\label{fig03}
\end{figure}

\begin{figure}
\centering
\rotatebox{-90}{\includegraphics[width=6.0cm]{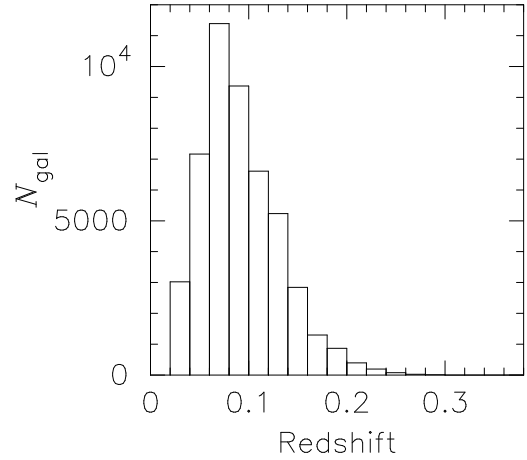}}
\caption{
   Frequency distribution of the redshift of the SDSS DR4 galaxies
   after our sample selection (sample C) described in \S\S2.2.
   Galaxies at $z<0.028$ are not included (see text).
}
\label{fig04}
\end{figure}

\begin{figure}
\centering
\rotatebox{-90}{\includegraphics[width=6.0cm]{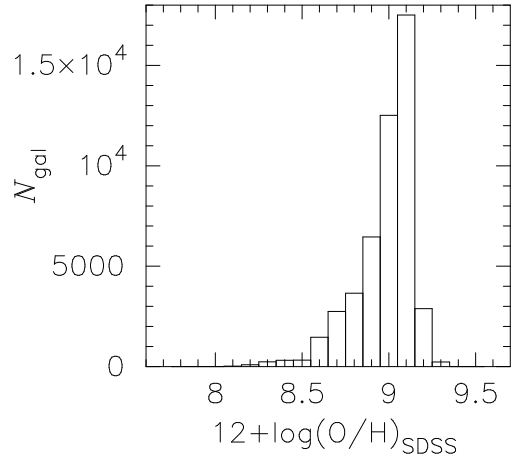}}
\caption{
   Frequency distribution of the oxygen abundance of galaxies
   in sample C.
}
\label{fig05}
\end{figure}

\clearpage

\begin{figure}
\centering
\rotatebox{-90}{\includegraphics[width=20.0cm]{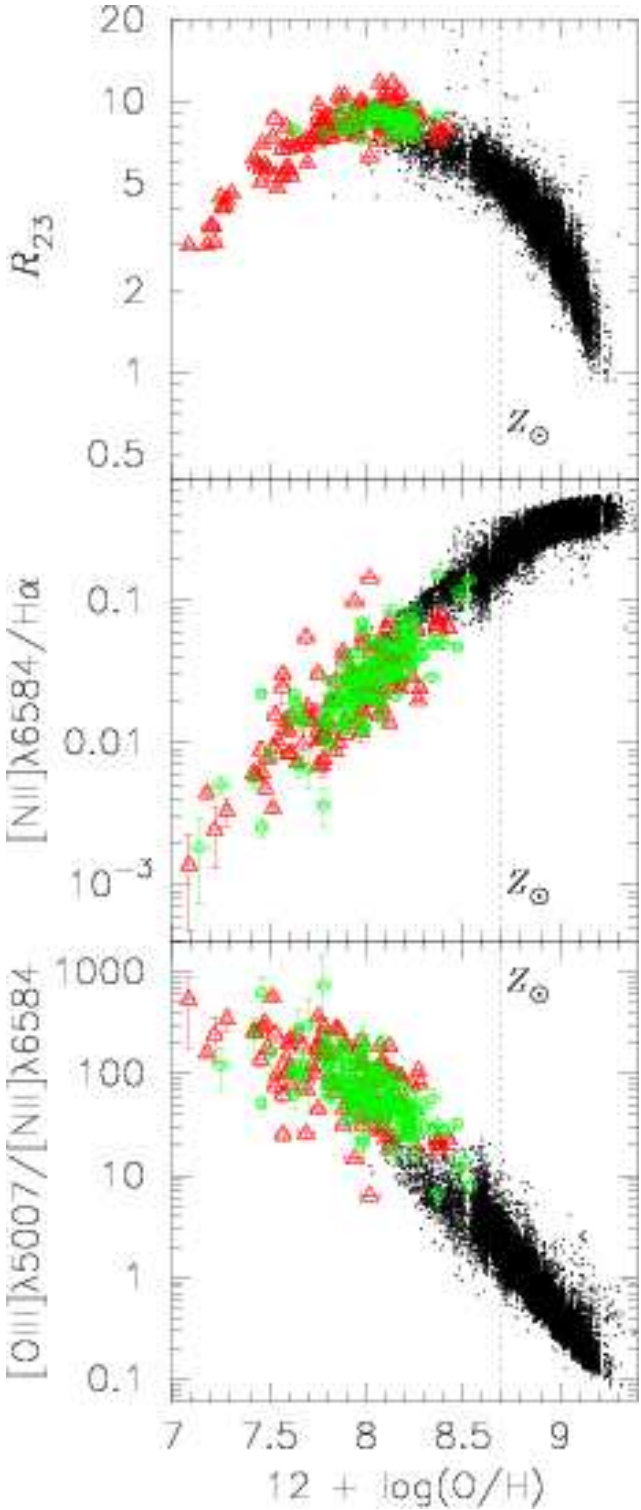}}
\caption{
   Emission-line flux ratios of $R_{23}$
   (= [$F$([O{\sc ii}]$\lambda$3727) + 
   1.327$\times$$F$([O{\sc iii}]$\lambda$5007)]
   $\! \! / \! $$F$(H$\beta$)), 
   $F$([N{\sc ii}]$\lambda$6584)/$F$(H$\alpha$), and 
   $F$([O{\sc iii}]$\lambda$5007)/$F$([N{\sc ii}]$\lambda$6584)
   for galaxies in sample A (red triangles), in sample B
   (green circles) and in sample C (black dots),
   as a function of the oxygen
   abundance. The compiled low-metallicity galaxies with an
   error of the oxygen abundance larger than 0.05 dex are not 
   plotted. Dotted lines denote the solar metallicity
   [12+log(O/H) = 8.69].
}
\label{fig06}
\end{figure}

\begin{figure}
\centering
\rotatebox{-90}{\includegraphics[width=20.0cm]{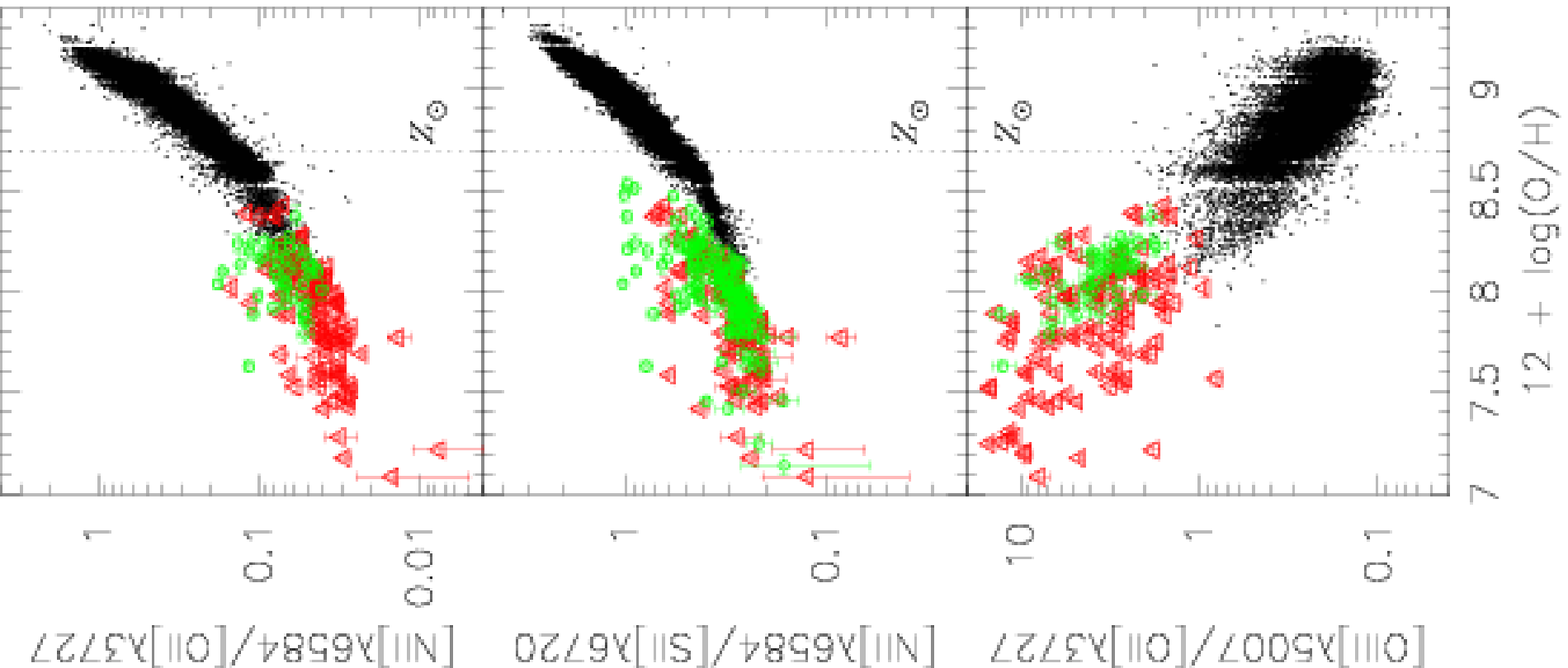}}
\caption{
   Same as Figure 6 but for the emission-line flux ratios of
   $F$([N{\sc ii}]$\lambda$6584)/$F$([O{\sc ii}]$\lambda$3727),
   $F$([N{\sc ii}]$\lambda$6584)/$F$([S{\sc ii}]$\lambda$6720),
   and $F$([O{\sc iii}]$\lambda$5007)/$F$([O{\sc ii}]$\lambda$3727).
}
\label{fig07}
\end{figure}

\clearpage

\begin{figure}
\centering
\rotatebox{-90}{\includegraphics[width=20.0cm]{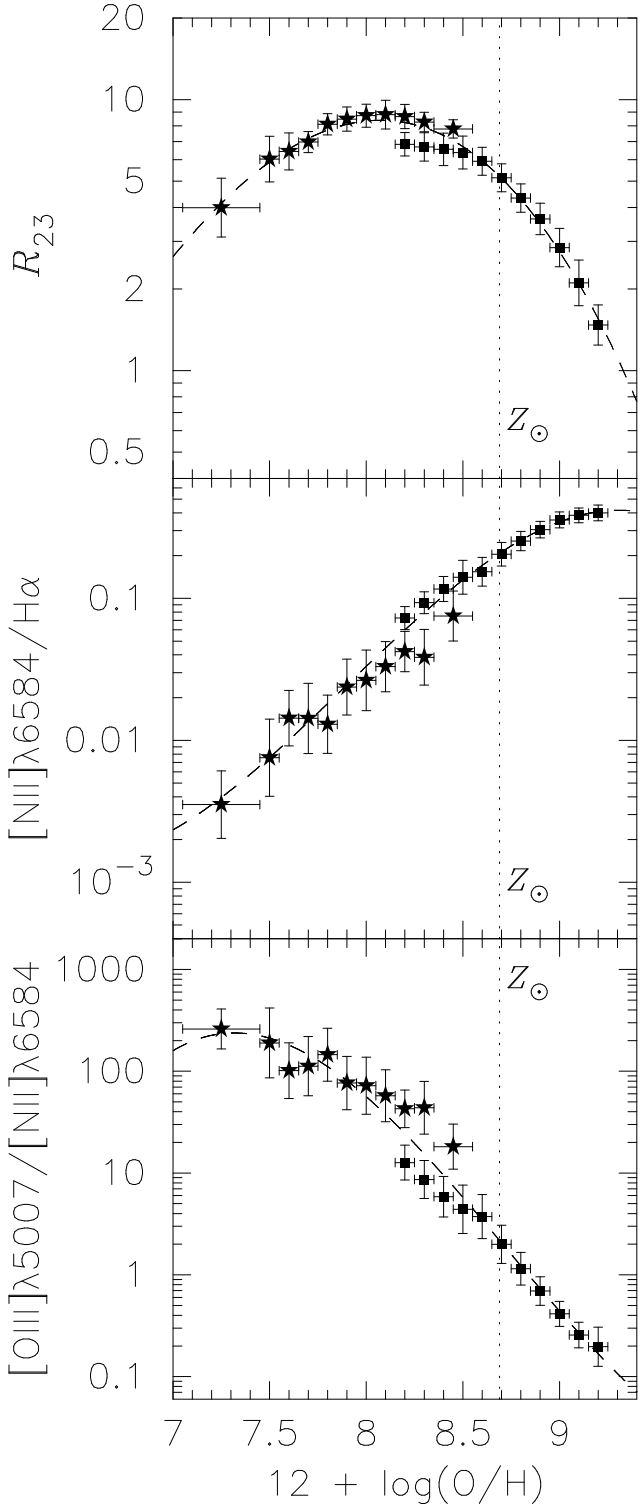}}
\caption{
   Same as Figure 6 but means and the RMS values are shown
   in each bin of oxygen abundance, instead of  
   individual data. Filled squares and filled stars denote
   the mean flux ratios for galaxies in sample C and those for
   galaxies in sample A+B, respectively.
   The errorbar denotes the RMS. The dashed line denotes
   the best-fit polynomial function, as described in the text.
   Dashed lines denote the solar metallicity [12+log(O/H) = 8.69].
}
\label{fig08}
\end{figure}

\begin{figure}
\centering
\rotatebox{-90}{\includegraphics[width=20.0cm]{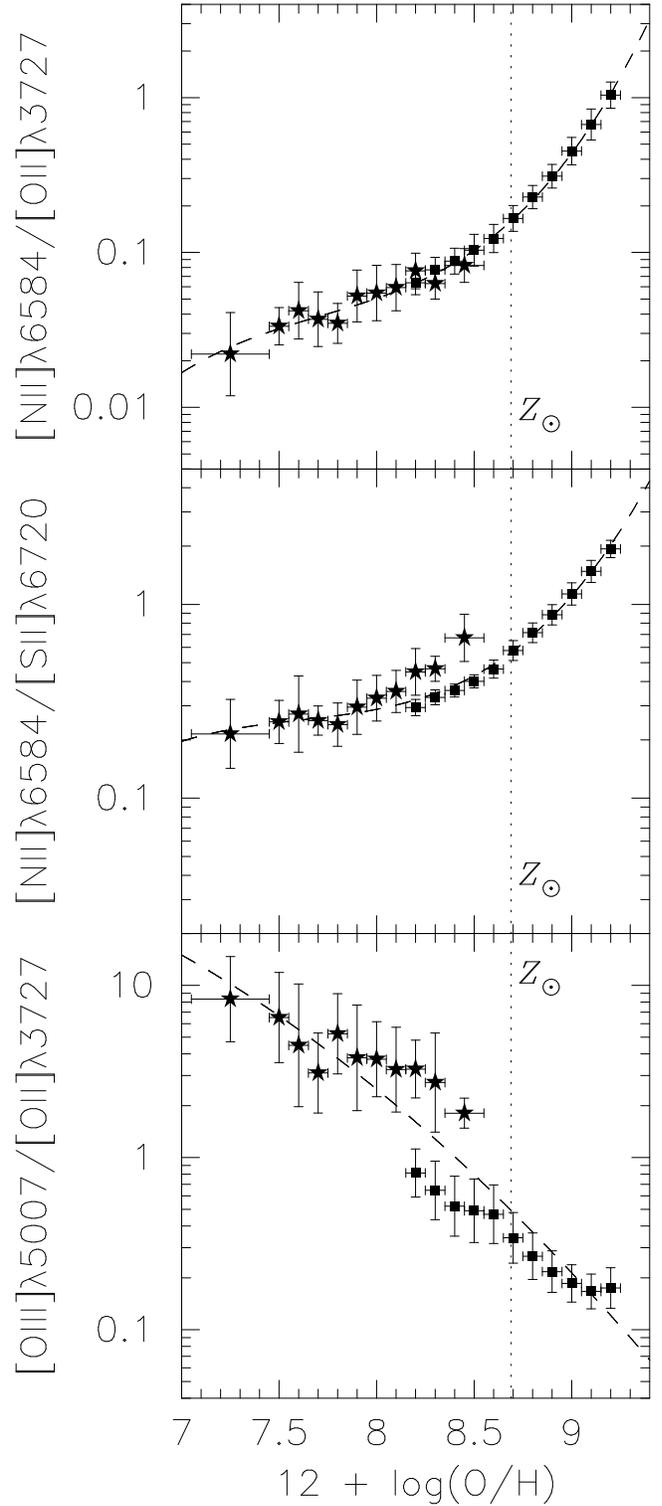}}
\caption{
   Same as Figure 8 but for the emission-line flux ratios of
   $F$([N{\sc ii}]$\lambda$6584)/$F$([O{\sc ii}]$\lambda$3727),
   $F$([N{\sc ii}]$\lambda$6584)/$F$([S{\sc ii}]$\lambda$6720),
   and $F$([O{\sc iii}]$\lambda$5007)/$F$([O{\sc ii}]$\lambda$3727).
}
\label{fig09}
\end{figure}

\clearpage

\begin{figure}
\centering
\rotatebox{-90}{\includegraphics[width=8cm]{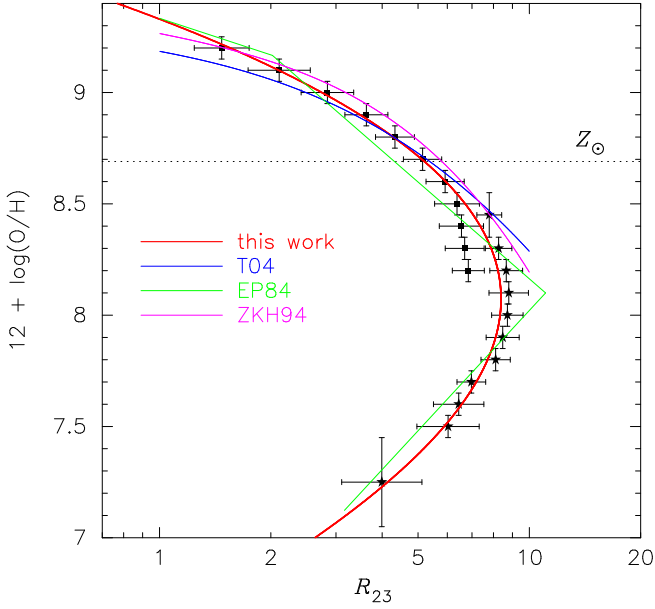}}
\caption{
   Comparison of our results with the previous
   empirical metallicity calibrations for the $R_{23}$ parameter.
   Solid red line denotes our calibration.
   Blue, green, and magenta lines denote the calibration
   given by Tremonti et al. (2004), Edmunds \& Pagel (1984), and
   Zaritsky et al. (1994), respectively.
   Symbols and errorbars are the same as those in Figure 8.
   Vertical dotted line denotes the solar metallicity 
   [12+log(O/H) = 8.69].
}
\label{fig10}
\end{figure}

\begin{figure}
\centering
%\vspace{10mm}
\rotatebox{-90}{\includegraphics[width=8cm]{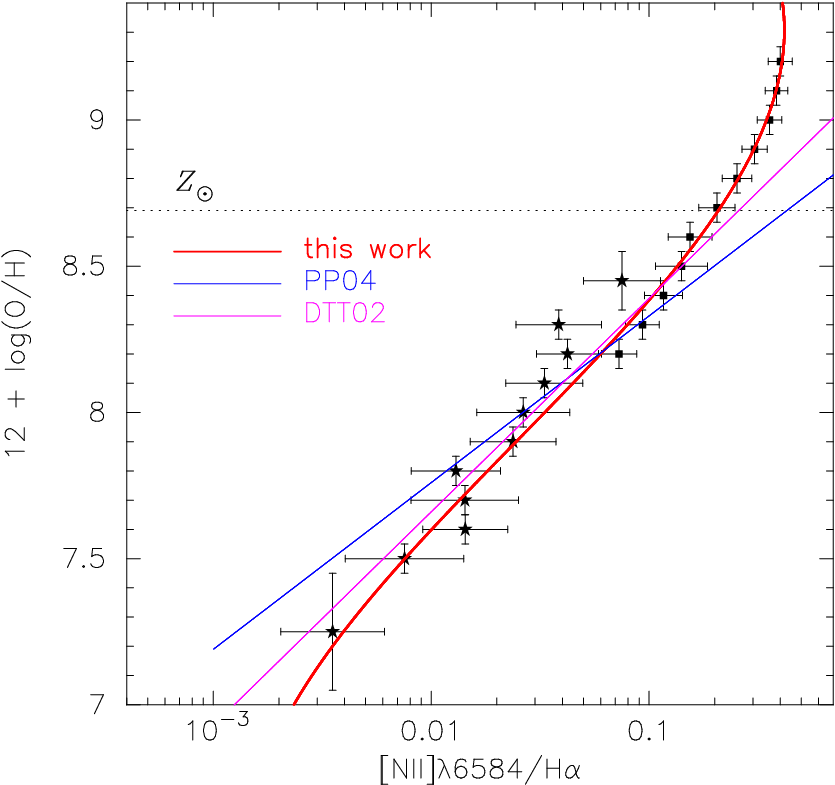}}
\caption{
   Comparison of our results with the previous
   empirical metallicity calibrations for 
   $F$([N{\sc ii}]$\lambda$6584)/$F$(H$\alpha$).
   Solid red line denotes our calibration.
   Blue and magenta lines denote the calibration
   given by Pettini \& Pagel (2004) and Denicol\'{o} et al. (2002),
   respectively.
   Symbols and errorbars are the same as those in Figure 8.
   Vertical dotted line denotes the solar metallicity 
   [12+log(O/H) = 8.69].
}
\label{fig11}
\end{figure}

\begin{figure}
\centering
\rotatebox{-90}{\includegraphics[width=8cm]{5216f12.eps}}
\caption{
   Comparison of our results with the previous
   empirical metallicity calibrations for 
   $F$([O{\sc iii}]$\lambda$5007)/$F$([N{\sc ii}]$\lambda$6584).
   Solid red line denotes our calibration,
   and blue line denotes the calibration
   given by Pettini \& Pagel (2004).
   Symbols and errorbars are the same as those in Figure 8.
   Vertical dotted line denotes the solar metallicity 
   [12+log(O/H) = 8.69].
}
\label{fig12}
\end{figure}

\clearpage

\begin{figure}
\centering
\rotatebox{-90}{\includegraphics[width=20cm]{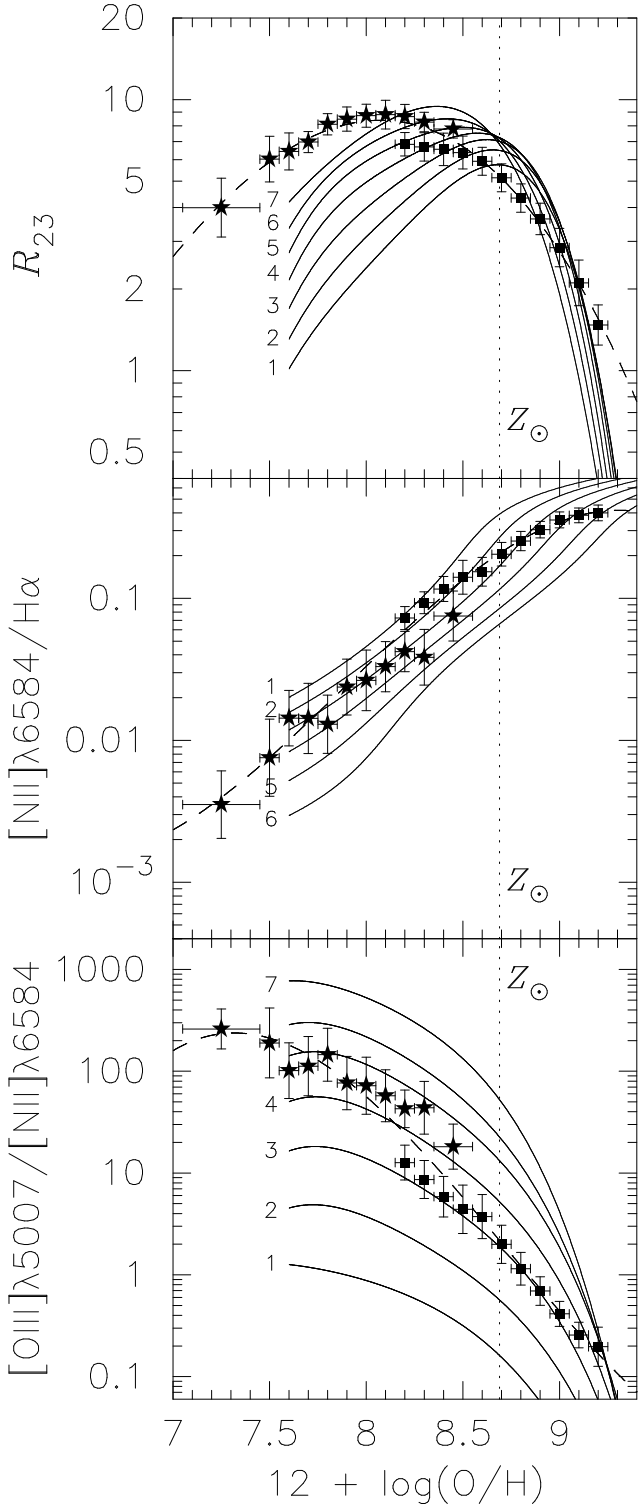}}
\caption{
   The averaged flux ratios and the best-fit polynomial 
   functions of the metallicity dependence of $R_{23}$,
   $F$([N{\sc ii}]$\lambda$6584)/$F$(H$\alpha$) and 
   $F$([O{\sc iii}]$\lambda$5007)/$F$([N{\sc ii}]$\lambda$6584)
   (dashed line) are compared with the predictions of
   photoionization models (solid lines: Kewley \& Dopita 2002; 
   Kobulnicky \& Kewley 2004). The lines with a digit
   1, 2, 3, 4, 5, 6, and 7 denote the model predictions with
   the ionization parameter of log $U$ = --3.8, --3.5,
   --3.2, --2.9, --2.6, --2.3, and --2.0, respectively.
   Dotted line denotes the solar metallicity [12+log(O/H) = 8.69].
}
\label{fig13}
\end{figure}

\begin{figure}
\centering
\rotatebox{-90}{\includegraphics[width=20cm]{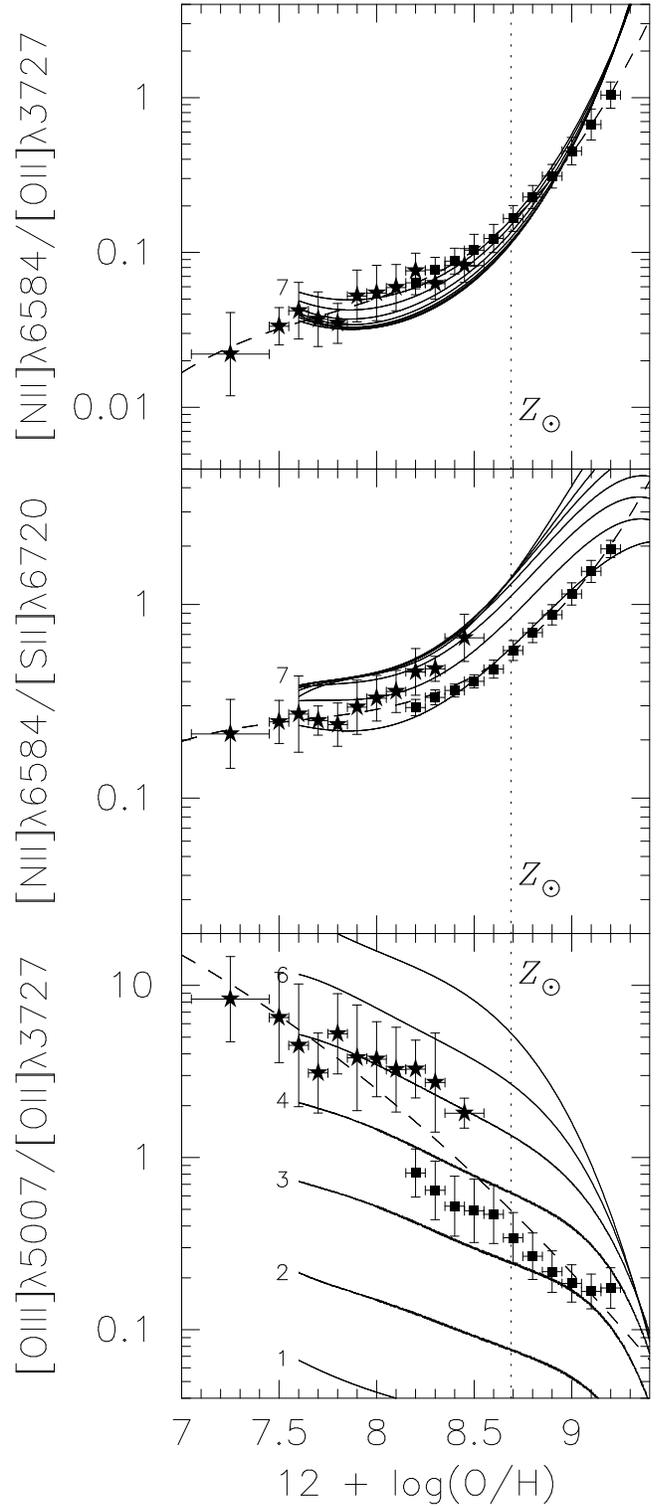}}
\caption{
   Same as Figure 13 but for the emission-line flux ratios of
   $F$([N{\sc ii}]$\lambda$6584)/$F$([O{\sc ii}]$\lambda$3727),
   $F$([N{\sc ii}]$\lambda$6584)/$F$([S{\sc ii}]$\lambda$6720),
   and $F$([O{\sc iii}]$\lambda$5007)/$F$([O{\sc ii}]$\lambda$3727).
}
\label{fig14}
\end{figure}

\clearpage

\begin{figure}
\centering
\rotatebox{-90}{\includegraphics[width=13.5cm]{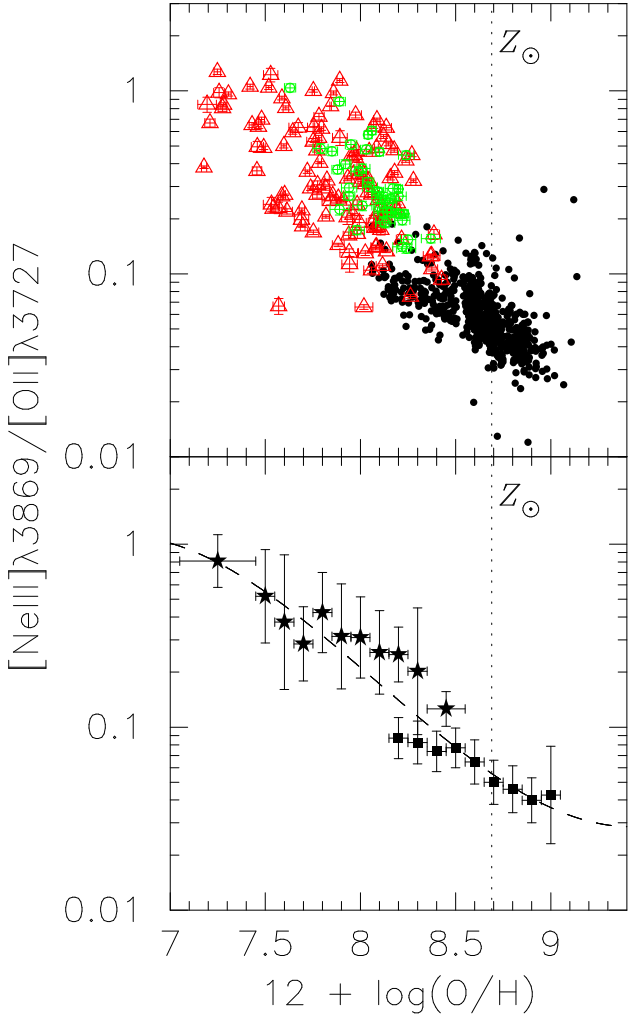}}
\caption{
   ($Upper$) Emission-line flux ratios of 
   $F$([Ne{\sc iii}]$\lambda$3869)/ $F$([O{\sc ii}]$\lambda$3727)
   of the galaxies in sample A (red triangles), those in 
   sample B (green circles), and those in sample C (black dots), 
   as a function of the oxygen abundance. As for sample C, only 
   the objects with S/N([Ne{\sc iii}]) $>$ 10 and 
   S/N([O{\sc ii}]) $>$ 10 (cataloged values) are plotted.
   ($Lower$) Same as the upper panel but the mean and the RMS
   values are shown for each bin of the oxygen abundance,
   instead of the individual data. Filled stars and filled
   squares denote the mean flux ratios for galaxies in sample A+B
   and those for galaxies in sample C, respectively.
   The errorbar denotes the RMS for each metallicity bin.
   The dashed line denotes the best-fit polynomial (second-order) 
   function.
}
\label{fig15}
\end{figure}

\begin{figure}
\centering
\rotatebox{-90}{\includegraphics[width=13.5cm]{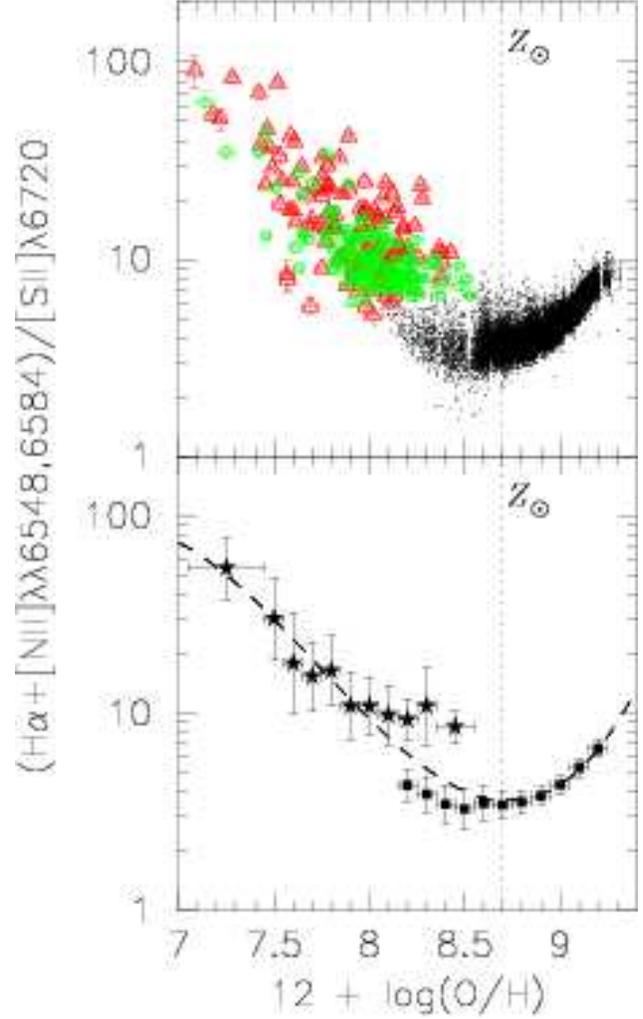}}
\caption{
   Same as Figure 15 but for the emission-line flux ratio of
   $F$(H$\alpha$+[N{\sc ii}]$\lambda\lambda$6548,6584)/$F$([S{\sc ii}]$\lambda$6720).
}
\label{fig16}
\end{figure}

\clearpage

\begin{figure}
\centering
\rotatebox{-90}{\includegraphics[width=13.5cm]{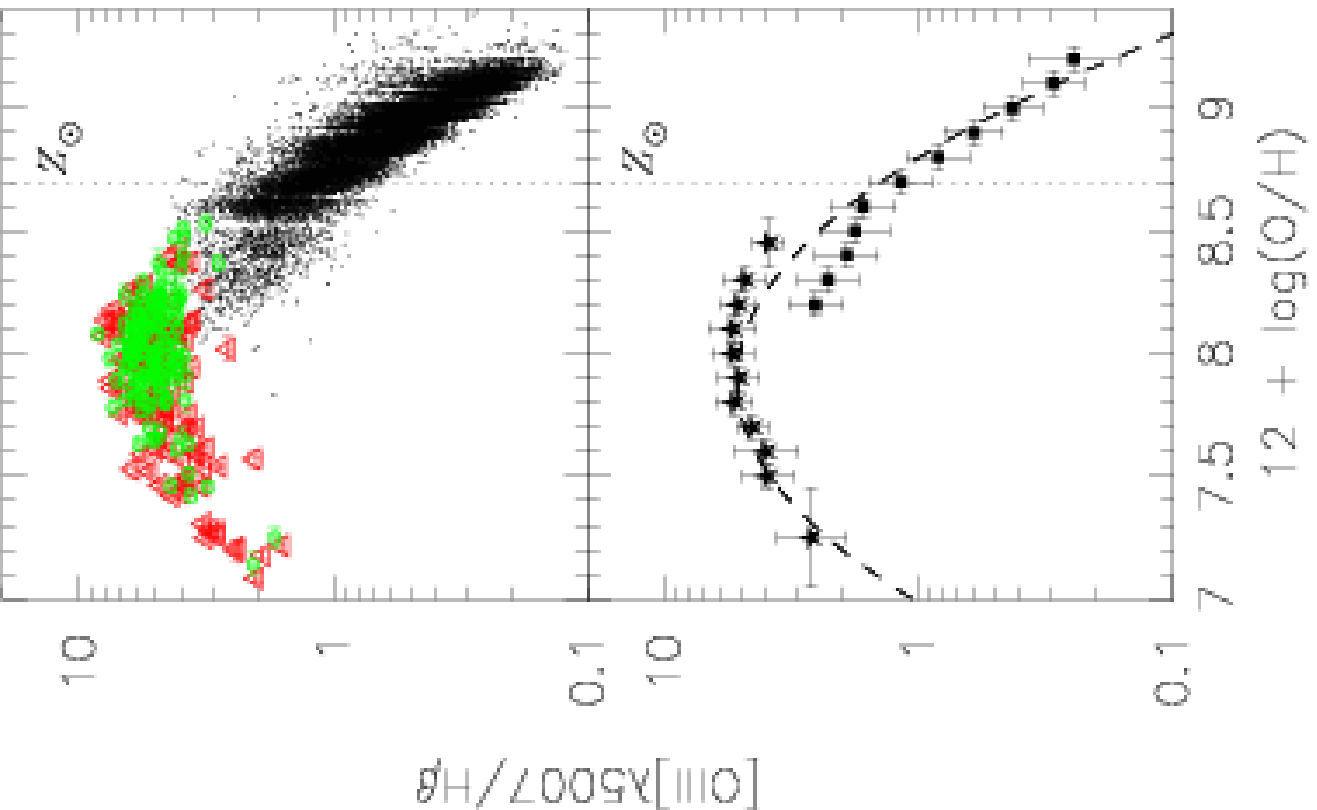}}
\caption{
   Same as Figure 15 but for the emission-line flux ratio of
   $F$([O{\sc iii}]$\lambda$5007)/$F$(H$\beta$).
}
\label{fig17}
\end{figure}

\begin{figure}
\centering
\rotatebox{-90}{\includegraphics[width=13.5cm]{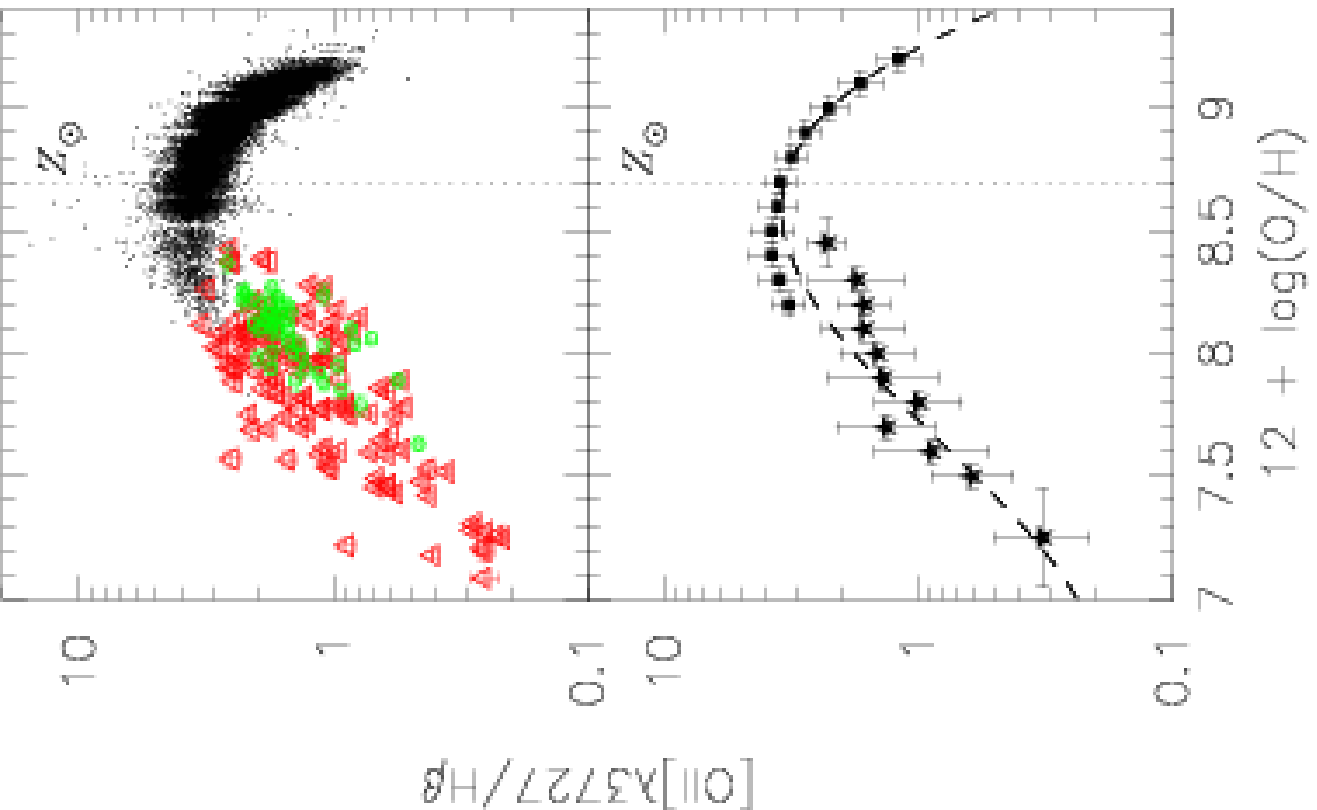}}
\caption{
   Same as Figure 15 but for the emission-line flux ratio of
   $F$([O{\sc ii}]$\lambda$3727)/$F$(H$\beta$).
}
\label{fig18}
\end{figure}

\clearpage

\begin{figure*}
\centering
\rotatebox{0}{\includegraphics[width=17.6cm]{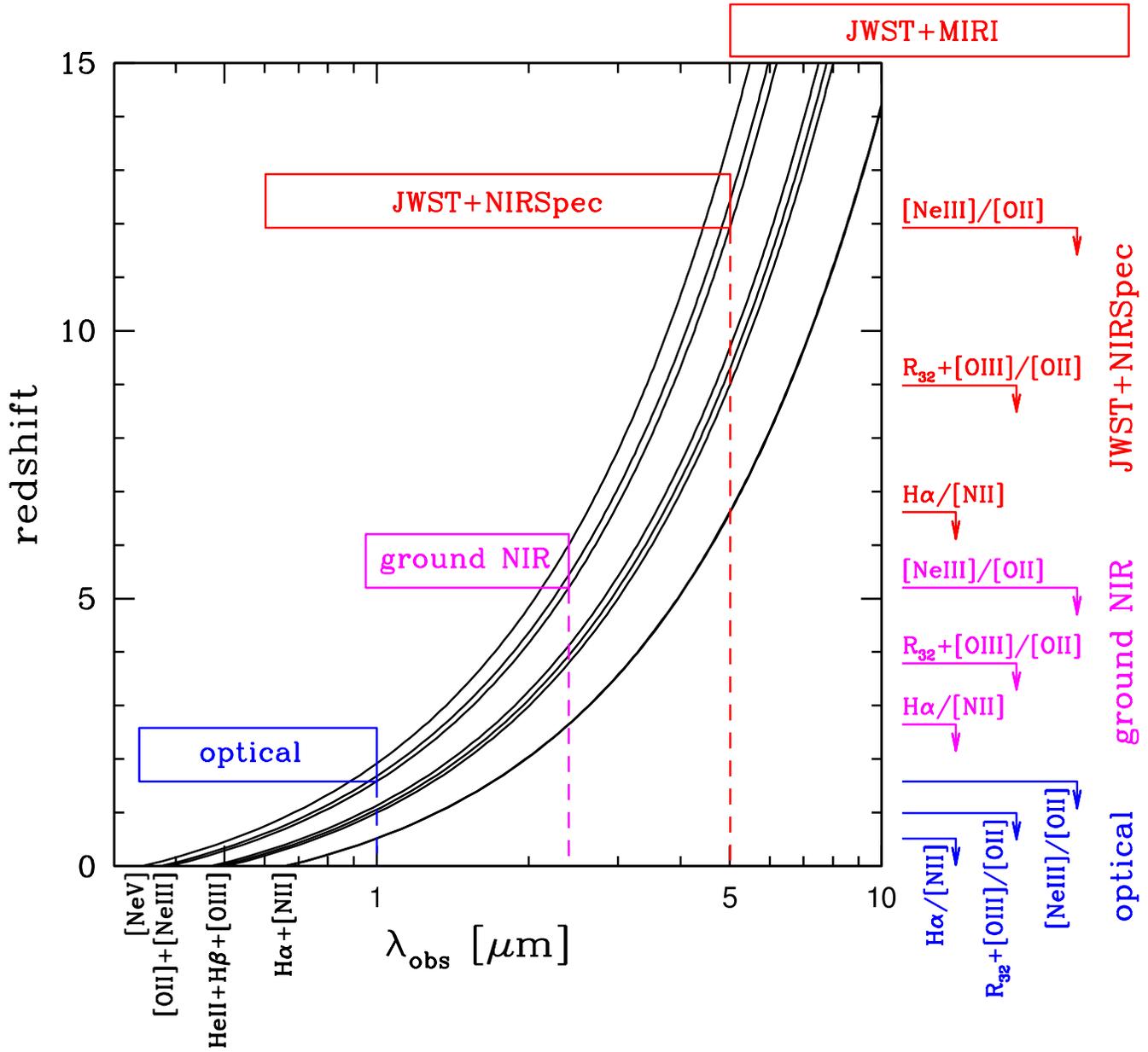}}
\caption{
  Schematic view of the availability of various metallicity 
  diagnostics for each redshift. The black solid curves
  indicate the effect of redshift for some of the diagnostic
  lines discussed in this paper. The colored boxes indicate
  the wavelength coverages of optical spectrometers (blue),
  of ground-based near-IR spectrometers (magenta), and of
  NIRSpec/MIRI on board of JWST (red). The marks on the
  right of the diagram indicate the maximum redshift at
  which some of the metallicity diagnostics can be used
  with the various facilities.
}
\label{fig19}
\end{figure*}

\begin{table*}
\centering
\caption{Re-calculated properties of the compiled 
         low-metallicity galaxies}
\label{table:01a}
\begin{tabular}{l c c c c c}
\hline\hline
\noalign{\smallskip}

Object                            & 
$R_{23}$                          &
$n_{\rm H}$(S$^+$)$^{\mathrm{a}}$ &
$t$(O$^{2+}$)$^{\mathrm{b}}$      &
12 + log(O/H)                     &
Ref.$^{\mathrm{c}}$            \\

\noalign{\smallskip}
\hline 
\noalign{\smallskip}

{} HS 0029+1748     &  8.624 $\pm$ 0.168 & $  70^{  +47}_{ -42}$ & $1.286 \pm 0.016$         & 8.046 $\pm$ 0.016 & I04b \\ 
{} HS 0111+2115     &  9.142 $\pm$ 0.179 & $<   22$              & $1.108^{+0.063}_{-0.066}$ & 8.290 $\pm$ 0.076 & I04b \\ 
{} HS 0122+0743     &  6.669 $\pm$ 0.127 & $  42^{  +59}_{ -42}$ & $1.777 \pm 0.025$         & 7.597 $\pm$ 0.014 & I04b \\ 
{} HS 0128+2832     & 10.272 $\pm$ 0.197 & $ 104^{  +40}_{ -37}$ & $1.256 \pm 0.011$         & 8.147 $\pm$ 0.013 & I04b \\ 
{} HS 0134+3415     & 10.485 $\pm$ 0.204 & $ 200^{  +76}_{ -68}$ & $1.639 \pm 0.019$         & 7.858 $\pm$ 0.014 & I04b \\ 
{} HS 0735+3512     &  9.640 $\pm$ 0.163 & $  75^{  +39}_{ -36}$ & $1.205 \pm 0.015$         & 8.193 $\pm$ 0.014 & I04b \\ 
{} HS 0811+4913     &  9.861 $\pm$ 0.184 & $  44^{  +51}_{ -44}$ & $1.449 \pm 0.016$         & 7.968 $\pm$ 0.013 & I04b \\ 
{} HS 0837+4717     &  8.061 $\pm$ 0.153 & $ 373^{ +117}_{-101}$ & $1.954 \pm 0.026$         & 7.587 $\pm$ 0.013 & I04b \\ 
{} HS 0924+3821     &  8.617 $\pm$ 0.152 & $  16^{  +44}_{ -16}$ & $1.256 \pm 0.021$         & 8.089 $\pm$ 0.018 & I04b \\ 
{} HS 1028+3843     & 10.464 $\pm$ 0.206 & $ 492^{ +138}_{-117}$ & $1.582 \pm 0.017$         & 7.891 $\pm$ 0.014 & I04b \\ 
{} HS 1213+3636A    &  7.353 $\pm$ 0.112 & $  35^{  +35}_{ -32}$ & $1.074^{+0.026}_{-0.027}$ & 8.263 $\pm$ 0.033 & I04b \\ 
{} HS 1214+3801     &  9.095 $\pm$ 0.162 & $  20^{  +37}_{ -20}$ & $1.339 \pm 0.013$         & 8.026 $\pm$ 0.012 & I04b \\ 
{} HS 1311+3628     &  8.570 $\pm$ 0.146 & $  95^{  +38}_{ -35}$ & $1.141 \pm 0.013$         & 8.199 $\pm$ 0.015 & I04b \\ 
{} HS 2236+1344     &  6.975 $\pm$ 0.135 & $  86^{ +131}_{ -86}$ & $2.123 \pm 0.032$         & 7.464 $\pm$ 0.013 & I04b \\ 
{} HS 2359+1659     &  9.608 $\pm$ 0.125 & $<   10$              & $1.189^{+0.016}_{-0.017}$ & 8.179 $\pm$ 0.017 & I04b \\ 
{} IC 0010 1        &  6.757 $\pm$ 0.400 & (10)$^d$              & $1.015^{+0.077}_{-0.085}$ & 8.301 $\pm$ 0.096 & L03b \\ 
{} IC 0010 2        &  6.607 $\pm$ 0.369 & (10)$^d$              & $0.984^{+0.077}_{-0.086}$ & 8.335 $\pm$ 0.097 & L03b \\ 
{} IC 0010 3        &  7.658 $\pm$ 0.624 & (10)$^d$              & $1.236^{+0.155}_{-0.169}$ & 8.083 $\pm$ 0.128 & L03b \\ 
{} IC 1613          &  7.641 $\pm$ 0.822 & $<  167$              & $1.796^{+0.161}_{-0.158}$ & 7.643 $\pm$ 0.085 & L03a \\ 
{} IC 5152          &  5.852 $\pm$ 0.509 & $  21^{ +275}_{ -21}$ & $1.240^{+0.166}_{-0.183}$ & 7.955 $\pm$ 0.135 & L03a \\ 
{} KISSB 0023       &  5.251 $\pm$ 0.238 & $<  172$              & $1.833^{+0.094}_{-0.093}$ & 7.570 $\pm$ 0.030 & M04  \\ 
{} KISSB 0061       &  7.348 $\pm$ 0.239 & $<   10$              & $1.565^{+0.037}_{-0.038}$ & 7.772 $\pm$ 0.024 & L04  \\ 
{} KISSB 0086       &  8.177 $\pm$ 0.242 & $<   10$              & $1.241 \pm 0.028$         & 8.088 $\pm$ 0.025 & L04  \\ 
{} KISSB 0171       &  8.276 $\pm$ 0.247 & $<   35$              & $1.189 \pm 0.020$         & 8.121 $\pm$ 0.022 & L04  \\ 
{} KISSB 0175       &  9.346 $\pm$ 0.295 & $ 166^{ +139}_{-110}$ & $1.344 \pm 0.024$         & 8.041 $\pm$ 0.021 & L04  \\ 
{} KISSR 0049       &  7.786 $\pm$ 0.398 & $  95^{ +291}_{ -95}$ & $1.305^{+0.083}_{-0.086}$ & 8.028 $\pm$ 0.059 & M04  \\ 
{} KISSR 0073       &  7.424 $\pm$ 0.382 & $  13^{ +194}_{ -13}$ & $1.366^{+0.056}_{-0.057}$ & 7.943 $\pm$ 0.042 & L04  \\ 
{} KISSR 0085       &  5.316 $\pm$ 0.222 & $ 833^{ +727}_{-412}$ & $1.775^{+0.104}_{-0.103}$ & 7.532 $\pm$ 0.043 & M04  \\ 
{} KISSR 0087       &  7.827 $\pm$ 0.247 & $  71^{ +111}_{ -71}$ & $0.996 \pm 0.023$         & 8.386 $\pm$ 0.030 & L04  \\ 
{} KISSR 0116       &  7.826 $\pm$ 0.241 & $<   90$              & $1.205^{+0.021}_{-0.022}$ & 8.107 $\pm$ 0.022 & L04  \\ 
{} KISSR 0286       &  7.710 $\pm$ 0.235 & $  29^{ +103}_{ -29}$ & $1.103 \pm 0.022$         & 8.216 $\pm$ 0.028 & L04  \\ 
{} KISSR 0310       &  9.501 $\pm$ 0.551 & $  26^{ +198}_{ -26}$ & $1.528 \pm 0.050$         & 7.892 $\pm$ 0.039 & L04  \\ 
{} KISSR 0311       &  8.935 $\pm$ 0.468 & $  32^{ +192}_{ -32}$ & $1.387^{+0.045}_{-0.046}$ & 7.994 $\pm$ 0.036 & L04  \\ 
{} KISSR 0396       &  8.023 $\pm$ 0.238 & $  27^{  +69}_{ -27}$ & $1.406^{+0.046}_{-0.047}$ & 7.938 $\pm$ 0.029 & M04  \\ 
{} KISSR 0666       &  8.630 $\pm$ 0.397 & $<   10$              & $2.153^{+0.092}_{-0.090}$ & 7.528 $\pm$ 0.038 & M04  \\ 
{} KISSR 0675       &  8.773 $\pm$ 0.583 & $1138^{+2735}_{-763}$ & $1.507^{+0.112}_{-0.114}$ & 7.890 $\pm$ 0.071 & M04  \\ 
{} KISSR 0814       &  8.620 $\pm$ 0.276 & $  79^{ +132}_{ -79}$ & $1.373 \pm 0.026$         & 7.983 $\pm$ 0.022 & L04  \\ 
{} KISSR 1013       &  7.271 $\pm$ 0.295 & $ 474^{ +346}_{-239}$ & $1.781^{+0.105}_{-0.104}$ & 7.690 $\pm$ 0.038 & M04  \\ 
{} KISSR 1194       &  8.836 $\pm$ 0.356 & $  51^{ +240}_{ -51}$ & $1.458^{+0.041}_{-0.042}$ & 7.930 $\pm$ 0.029 & M04  \\ 
{} KISSR 1490       &  6.676 $\pm$ 0.308 & $<  381$              & $1.903^{+0.110}_{-0.109}$ & 7.568 $\pm$ 0.040 & M04  \\ 
{} KISSR 1778       &  6.628 $\pm$ 0.325 & $  69^{ +383}_{ -69}$ & $1.305^{+0.091}_{-0.094}$ & 7.955 $\pm$ 0.064 & M04  \\ 
{} KISSR 1845       &  9.440 $\pm$ 0.381 & $  50^{ +240}_{ -50}$ & $1.325 \pm 0.033$         & 8.069 $\pm$ 0.028 & M04  \\ 
{} Mrk 0005         &  7.192 $\pm$ 0.106 & $  13^{  +54}_{ -13}$ & $1.220^{+0.051}_{-0.052}$ & 8.058 $\pm$ 0.041 & I98  \\ 
{} Mrk 0022         &  8.726 $\pm$ 0.101 & $  71^{  +55}_{ -49}$ & $1.349^{+0.020}_{-0.021}$ & 8.002 $\pm$ 0.015 & I94  \\ 
{} Mrk 0035         &  7.892 $\pm$ 0.121 & $ 189^{  +36}_{ -34}$ & $1.021^{+0.012}_{-0.013}$ & 8.368 $\pm$ 0.015 & I04b \\ 
{} Mrk 0036         &  7.708 $\pm$ 0.092 & $<  102$              & $1.524 \pm 0.037$         & 7.816 $\pm$ 0.021 & I98  \\ 
{} Mrk 0067         &  9.309 $\pm$ 0.176 & $<   10$              & $1.320 \pm 0.024$         & 8.059 $\pm$ 0.019 & I04b \\ 
{} Mrk 0116         &  2.937 $\pm$ 0.017 & $  86^{ +112}_{ -86}$ & $1.927 \pm 0.038$         & 7.178 $\pm$ 0.014 & I97  \\ 
{} Mrk 0116 1       &  2.935 $\pm$ 0.043 & $< 1369$              & $2.133^{+0.064}_{-0.063}$ & 7.084 $\pm$ 0.023 & P92  \\ 
{} Mrk 0116 2       &  3.009 $\pm$ 0.071 & $  68^{ +722}_{ -68}$ & $1.979^{+0.092}_{-0.090}$ & 7.217 $\pm$ 0.027 & P92  \\ 
{} Mrk 0162         &  8.180 $\pm$ 0.083 & $<   10$              & $1.194^{+0.043}_{-0.044}$ & 8.116 $\pm$ 0.034 & I98  \\ 
{} Mrk 0178         &  8.517 $\pm$ 0.245 & $ 122^{ +234}_{-122}$ & $1.588 \pm 0.104$         & 7.816 $\pm$ 0.057 & G00  \\ 
{} Mrk 0193         &  8.906 $\pm$ 0.101 & $ 172^{  +93}_{ -81}$ & $1.639 \pm 0.019$         & 7.795 $\pm$ 0.011 & I94  \\ 
{} Mrk 0209         &  8.075 $\pm$ 0.018 & $  46^{  +45}_{ -42}$ & $1.630 \pm 0.007$         & 7.755 $\pm$ 0.004 & I97  \\ 
{} Mrk 0450 1       &  8.514 $\pm$ 0.144 & $ 132^{  +39}_{ -36}$ & $1.173 \pm 0.013$         & 8.154 $\pm$ 0.014 & I04b \\ 
{} Mrk 0450 2       &  8.661 $\pm$ 0.166 & $<   21$              & $1.251^{+0.028}_{-0.029}$ & 8.094 $\pm$ 0.024 & I04b \\ 
{} Mrk 0475         &  8.392 $\pm$ 0.111 & $<   45$              & $1.411 \pm 0.028$         & 7.933 $\pm$ 0.019 & I94  \\ 

\noalign{\smallskip}
\hline
\end{tabular}
\end{table*}

\begin{table*}
\addtocounter{table}{-1}
\caption{Re-calculated properties of the compiled low-metallicity galaxies 
         (continued)}
\label{table:01b}
\centering
\begin{tabular}{l c c c c c}
\hline\hline
\noalign{\smallskip}

Object                            & 
$R_{23}$                          &
$n_{\rm H}$(S$^+$)$^{\mathrm{a}}$ &
$t$(O$^{2+}$)$^{\mathrm{b}}$      &
12 + log(O/H)                     &
Ref.$^{\mathrm{c}}$           \\

\noalign{\smallskip}
\hline 

{} Mrk 0487         &  8.413 $\pm$ 0.157 & $  63^{  +92}_{ -63}$ & $1.266^{+0.057}_{-0.058}$ & 8.076 $\pm$ 0.043 & I97  \\ 
{} Mrk 0600         &  8.578 $\pm$ 0.101 & $  58^{  +41}_{ -38}$ & $1.579 \pm 0.020$         & 7.824 $\pm$ 0.012 & I98  \\ 
{} Mrk 0724         &  8.618 $\pm$ 0.149 & $  19^{  +37}_{ -19}$ & $1.296^{+0.014}_{-0.015}$ & 8.045 $\pm$ 0.013 & I04b \\ 
{} Mrk 0750         &  8.357 $\pm$ 0.079 & $<   10$              & $1.205^{+0.023}_{-0.024}$ & 8.128 $\pm$ 0.021 & I98  \\ 
{} Mrk 0930         &  7.905 $\pm$ 0.084 & $  56^{  +40}_{ -36}$ & $1.236^{+0.037}_{-0.038}$ & 8.084 $\pm$ 0.029 & I98  \\ 
{} Mrk 1063         &  6.183 $\pm$ 0.106 & $  96^{  +43}_{ -39}$ & $1.027^{+0.058}_{-0.062}$ & 8.260 $\pm$ 0.069 & I04b \\ 
{} Mrk 1089         &  5.554 $\pm$ 0.057 & $  92^{  +43}_{ -39}$ & $1.108^{+0.069}_{-0.074}$ & 8.090 $\pm$ 0.070 & I98  \\ 
{} Mrk 1236         &  9.550 $\pm$ 0.170 & $  47^{  +35}_{ -33}$ & $1.225 \pm 0.012$         & 8.157 $\pm$ 0.013 & I04b \\ 
{} Mrk 1271         &  9.680 $\pm$ 0.074 & $  52^{  +51}_{ -47}$ & $1.411 \pm 0.018$         & 7.996 $\pm$ 0.012 & I98  \\ 
{} Mrk 1315         &  9.104 $\pm$ 0.164 & $  11^{  +30}_{ -11}$ & $1.103 \pm 0.009$         & 8.270 $\pm$ 0.013 & I04b \\ 
{} Mrk 1328         &  6.981 $\pm$ 0.165 & $  25^{  +83}_{ -25}$ & $0.937^{+0.099}_{-0.119}$ & 8.457 $\pm$ 0.106 & V03  \\ 
{} Mrk 1329         &  8.539 $\pm$ 0.150 & $  18^{  +31}_{ -18}$ & $1.080 \pm 0.009$         & 8.278 $\pm$ 0.013 & I04b \\ 
{} Mrk 1409         &  8.754 $\pm$ 0.138 & $ 599^{ +124}_{-106}$ & $1.362^{+0.066}_{-0.067}$ & 8.025 $\pm$ 0.040 & I97  \\ 
{} Mrk 1416         &  8.098 $\pm$ 0.065 & $<   10$              & $1.514 \pm 0.031$         & 7.854 $\pm$ 0.016 & I97  \\ 
{} Mrk 1434         &  7.640 $\pm$ 0.049 & $<   10$              & $1.551 \pm 0.015$         & 7.786 $\pm$ 0.009 & I97  \\ 
{} Mrk 1450         &  7.669 $\pm$ 0.052 & $<   43$              & $1.330 \pm 0.016$         & 7.963 $\pm$ 0.012 & I94  \\ 
{} Mrk 1486         &  8.140 $\pm$ 0.059 & $  27^{  +40}_{ -27}$ & $1.468 \pm 0.022$         & 7.884 $\pm$ 0.013 & I97  \\ 
{} NGC 2363 A       &  9.358 $\pm$ 0.020 & $  85^{  +58}_{ -53}$ & $1.584 \pm 0.006$         & 7.843 $\pm$ 0.004 & I97  \\ 
{} NGC 2363 B       &  7.286 $\pm$ 0.074 & $  14^{  +70}_{ -14}$ & $1.496^{+0.033}_{-0.034}$ & 7.818 $\pm$ 0.019 & I97  \\ 
{} NGC 3109         &  6.221 $\pm$ 0.437 & (10)$^d$              & $1.463^{+0.262}_{-0.277}$ & 7.792 $\pm$ 0.138 & L03b \\ 
{} NGC 4214 A6      &  6.898 $\pm$ 0.162 & $  33^{  +93}_{ -33}$ & $1.051^{+0.047}_{-0.050}$ & 8.280 $\pm$ 0.062 & K96  \\ 
{} NGC 4214 C6      &  7.913 $\pm$ 0.176 & $  74^{  +98}_{ -74}$ & $0.983^{+0.024}_{-0.025}$ & 8.426 $\pm$ 0.029 & K96  \\ 
{} NGC 4861         &  8.801 $\pm$ 0.020 & $  74^{  +27}_{ -26}$ & $1.363 \pm 0.006$         & 7.987 $\pm$ 0.005 & I97  \\ 
{} PGC 18096        & 10.671 $\pm$ 0.124 & $ 195^{  +64}_{ -57}$ & $1.339 \pm 0.021$         & 8.086 $\pm$ 0.017 & G00  \\ 
{} PGC 27864 1      &  8.410 $\pm$ 0.151 & $ 116^{  +47}_{ -43}$ & $1.648 \pm 0.018$         & 7.771 $\pm$ 0.012 & I04b \\ 
{} PGC 27864 2      &  7.616 $\pm$ 0.135 & $<   10$              & $1.657 \pm 0.027$         & 7.738 $\pm$ 0.014 & I04b \\ 
{} PGC 37727        &  8.082 $\pm$ 0.129 & $  57^{  +39}_{ -36}$ & $1.256^{+0.029}_{-0.030}$ & 8.079 $\pm$ 0.022 & I04b \\ 
{} PGC 39188        &  7.325 $\pm$ 0.091 & $1088^{ +120}_{-107}$ & $1.014^{+0.026}_{-0.027}$ & 8.375 $\pm$ 0.030 & V03  \\ 
{} PGC 39402        &  7.368 $\pm$ 0.146 & $ 261^{ +117}_{-100}$ & $2.002 \pm 0.028$         & 7.518 $\pm$ 0.014 & I04  \\ 
{} PGC 39845        &  5.990 $\pm$ 0.181 & $ 140^{ +305}_{-140}$ & $1.689 \pm 0.146$         & 7.653 $\pm$ 0.051 & V03  \\ 
{} PGC 40521        &  5.549 $\pm$ 0.350 & $  57^{ +326}_{ -57}$ & $1.330^{+0.221}_{-0.241}$ & 7.864 $\pm$ 0.145 & V03  \\ 
{} PGC 40582 1      &  5.814 $\pm$ 0.113 & $ 108^{  +74}_{ -66}$ & $1.880 \pm 0.031$         & 7.480 $\pm$ 0.015 & I04a \\ 
{} PGC 40582 2      &  5.033 $\pm$ 0.103 & $<   10$              & $1.829^{+0.057}_{-0.056}$ & 7.456 $\pm$ 0.025 & I04a \\ 
{} PGC 40582 3      &  4.901 $\pm$ 0.187 & $<   62$              & $1.908^{+0.342}_{-0.327}$ & 7.410 $\pm$ 0.125 & I04a \\ 
{} PGC 40582 4      &  5.670 $\pm$ 0.131 & (10)$^d$              & $1.918^{+0.078}_{-0.077}$ & 7.458 $\pm$ 0.032 & I04a \\ 
{} PGC 40604        &  7.671 $\pm$ 0.266 & $<  127$              & $1.266^{+0.162}_{-0.176}$ & 8.058 $\pm$ 0.120 & V03  \\ 
{} PGC 40604 a      &  6.668 $\pm$ 0.279 & $ 155^{ +273}_{-155}$ & $1.275^{+0.157}_{-0.170}$ & 7.972 $\pm$ 0.118 & V03  \\ 
{} PGC 41360        &  7.255 $\pm$ 0.187 & $  39^{ +107}_{ -39}$ & $1.542 \pm 0.047$         & 7.775 $\pm$ 0.029 & V03  \\ 
{} PGC 42160        &  5.418 $\pm$ 0.238 & $<  123$              & $1.486^{+0.229}_{-0.238}$ & 7.744 $\pm$ 0.104 & V03  \\ 
{} PGC 49050        &  7.773 $\pm$ 0.538 & $<  157$              & $1.103^{+0.127}_{-0.144}$ & 8.250 $\pm$ 0.194 & L03a \\ 
{} SBS 0335--052     & 4.343 $\pm$ 0.041 & $ 275^{ +225}_{-172}$ & $2.040 \pm 0.036$         & 7.280 $\pm$ 0.014 & I98  \\ 
{} SBS 0335--052 E3  & 4.588 $\pm$ 0.090 & (10)$^d$              & $2.027 \pm 0.029$         & 7.306 $\pm$ 0.014 & P06  \\
{} SBS 0335--052 E4-5& 4.443 $\pm$ 0.093 & (10)$^d$              & $2.128^{+0.041}_{-0.040}$ & 7.248 $\pm$ 0.017 & P06  \\
{} SBS 0335--052 E7  & 3.457 $\pm$ 0.074 & (10)$^d$              & $1.974^{+0.063}_{-0.062}$ & 7.209 $\pm$ 0.027 & P06  \\
{} SBS 0335--052 E-NW& 3.446 $\pm$ 0.097 & (10)$^d$              & $2.001^{+0.125}_{-0.122}$ & 7.194 $\pm$ 0.049 & P06  \\
{} SBS 0335--052 E-SE& 4.087 $\pm$ 0.085 & (10)$^d$              & $1.979 \pm 0.053$         & 7.275 $\pm$ 0.023 & P06  \\
{} SBS 0335--052 W   & 2.550 $\pm$ 0.072 & (10)$^d$              & $1.974^{+0.260}_{-0.249}$ & 7.133 $\pm$ 0.073 & P06  \\ 
{} SBS 0749+568     &  8.144 $\pm$ 0.227 & $<   10$              & $1.528 \pm 0.081$         & 7.843 $\pm$ 0.045 & I97  \\ 
{} SBS 0749+582     & 11.727 $\pm$ 0.272 & $ 117^{ +155}_{-117}$ & $1.334 \pm 0.034$         & 8.135 $\pm$ 0.028 & I97  \\ 
{} SBS 0907+543     & 10.014 $\pm$ 0.241 & $ 118^{ +277}_{-118}$ & $1.444 \pm 0.043$         & 7.974 $\pm$ 0.031 & I97  \\ 
{} SBS 0926+606     &  8.117 $\pm$ 0.059 & $ 188^{  +48}_{ -45}$ & $1.434 \pm 0.023$         & 7.911 $\pm$ 0.014 & I97  \\ 
{} SBS 0940+544     &  5.853 $\pm$ 0.081 & $ 188^{ +187}_{-146}$ & $2.016 \pm 0.038$         & 7.430 $\pm$ 0.015 & I97  \\ 
{} SBS 0943+561     &  9.024 $\pm$ 0.424 & $ 271^{ +630}_{-271}$ & $1.758^{+0.130}_{-0.129}$ & 7.749 $\pm$ 0.060 & I97  \\ 
{} SBS 0948+532     &  8.841 $\pm$ 0.144 & $  73^{  +85}_{ -73}$ & $1.339 \pm 0.027$         & 8.014 $\pm$ 0.021 & I94  \\ 
{} SBS 1054+365     &  8.959 $\pm$ 0.090 & $<   27$              & $1.383 \pm 0.019$         & 7.978 $\pm$ 0.014 & I97  \\ 
{} SBS 1116+583B    &  7.014 $\pm$ 0.249 & $ 593^{ +617}_{-351}$ & $1.670 \pm 0.089$         & 7.673 $\pm$ 0.049 & I97  \\ 
{} SBS 1128+573     &  8.570 $\pm$ 0.212 & $ 211^{ +339}_{-211}$ & $1.689 \pm 0.062$         & 7.751 $\pm$ 0.033 & I97  \\ 

\noalign{\smallskip}
\hline
\end{tabular}
\end{table*}

\begin{table*}
\addtocounter{table}{-1}
\caption{Re-calculated properties of the compiled low-metallicity galaxies 
         (continued)}
\label{table:01c}
\centering
\begin{tabular}{l c c c c c}
\hline\hline
\noalign{\smallskip}

Object                            & 
$R_{23}$                          &
$n_{\rm H}$(S$^+$)$^{\mathrm{a}}$ &
$t$(O$^{2+}$)$^{\mathrm{b}}$      &
12 + log(O/H)                     &
Ref.$^{\mathrm{c}}$           \\

\noalign{\smallskip}
\hline 
\noalign{\smallskip}

{} SBS 1129+576a    &  3.851 $\pm$ 0.111 & $<  248$              & $1.899^{+0.271}_{-0.262}$ & 7.369 $\pm$ 0.069 & G03a \\ 
{} SBS 1129+576b    &  5.804 $\pm$ 0.247 & $<  287$              & $2.094^{+0.291}_{-0.275}$ & 7.475 $\pm$ 0.064 & G03a \\ 
{} SBS 1159+545     &  5.620 $\pm$ 0.051 & $  57^{  +54}_{ -50}$ & $1.852 \pm 0.020$         & 7.491 $\pm$ 0.009 & I98  \\ 
{} SBS 1205+557     &  7.067 $\pm$ 0.103 & $<   76$              & $1.607 \pm 0.067$         & 7.752 $\pm$ 0.029 & I97  \\ 
{} SBS 1211+540     &  6.814 $\pm$ 0.073 & $ 168^{ +124}_{-103}$ & $1.699 \pm 0.024$         & 7.644 $\pm$ 0.013 & I94  \\ 
{} SBS 1222+614     &  9.071 $\pm$ 0.050 & $  22^{  +28}_{ -22}$ & $1.425 \pm 0.012$         & 7.951 $\pm$ 0.009 & I97  \\ 
{} SBS 1249+493     &  7.359 $\pm$ 0.087 & $<   10$              & $1.648 \pm 0.024$         & 7.721 $\pm$ 0.012 & I98  \\ 
{} SBS 1319+579 A   &  9.913 $\pm$ 0.066 & $ 145^{  +26}_{ -25}$ & $1.310^{+0.011}_{-0.012}$ & 8.084 $\pm$ 0.010 & I97  \\ 
{} SBS 1319+579 B   &  6.797 $\pm$ 0.189 & $  40^{ +113}_{ -40}$ & $1.359^{+0.165}_{-0.175}$ & 7.911 $\pm$ 0.102 & I97  \\ 
{} SBS 1319+579 C   &  7.054 $\pm$ 0.061 & $  20^{  +33}_{ -20}$ & $1.136^{+0.035}_{-0.036}$ & 8.136 $\pm$ 0.032 & I97  \\ 
{} SBS 1331+493     &  8.129 $\pm$ 0.110 & $ 164^{  +93}_{ -81}$ & $1.602 \pm 0.027$         & 7.780 $\pm$ 0.016 & I94  \\ 
{} SBS 1331+493S    &  6.308 $\pm$ 0.138 & $<   79$              & $1.353^{+0.086}_{-0.088}$ & 7.885 $\pm$ 0.056 & T95  \\ 
{} SBS 1415+437     &  5.677 $\pm$ 0.025 & $  65^{  +31}_{ -30}$ & $1.703 \pm 0.011$         & 7.586 $\pm$ 0.005 & I98  \\ 
{} SBS 1415+437e1   &  5.649 $\pm$ 0.025 & $  48^{  +31}_{ -30}$ & $1.657 \pm 0.010$         & 7.601 $\pm$ 0.005 & G03c \\ 
{} SBS 1415+437e2   &  5.290 $\pm$ 0.079 & $  79^{ +114}_{ -79}$ & $1.597 \pm 0.056$         & 7.614 $\pm$ 0.028 & G03c \\ 
{} SBS 1420+544     &  9.683 $\pm$ 0.070 & $<   10$              & $1.764 \pm 0.011$         & 7.752 $\pm$ 0.006 & I98  \\ 
{} SBS 1533+469     &  8.690 $\pm$ 0.226 & $  46^{ +106}_{ -46}$ & $1.383^{+0.054}_{-0.055}$ & 7.984 $\pm$ 0.035 & T95  \\ 
{} SBS 1533+574 A   &  7.507 $\pm$ 0.081 & $  30^{  +48}_{ -30}$ & $1.444 \pm 0.057$         & 7.883 $\pm$ 0.031 & I97  \\ 
{} SBS 1533+574 B   &  9.107 $\pm$ 0.083 & $<   23$              & $1.246 \pm 0.029$         & 8.124 $\pm$ 0.023 & I97  \\ 
{} SDSS J0113+0052  &  3.402 $\pm$ 0.211 & (10)$^d$              & $2.317^{+0.325}_{-0.301}$ & 7.163 $\pm$ 0.076 & I06  \\
{} SDSS J0519+0007  &  6.176 $\pm$ 0.122 & $ 373^{ +308}_{-220}$ & $2.078 \pm 0.036$         & 7.420 $\pm$ 0.015 & I04b \\ 
{} SDSS J2104--0035 N& 4.044 $\pm$ 0.093 & (10)$^d$              & $2.008 \pm 0.066$         & 7.257 $\pm$ 0.028 & I06  \\
{} UGC 4305 5       &  5.383 $\pm$ 0.354 & (10)$^d$              & $1.607^{+0.147}_{-0.148}$ & 7.647 $\pm$ 0.064 & L03b \\ 
{} UGC 4305 7       &  5.150 $\pm$ 0.280 & (10)$^d$              & $1.354^{+0.208}_{-0.224}$ & 7.806 $\pm$ 0.129 & L03b \\ 
{} UGC 4305 8       &  5.010 $\pm$ 0.306 & (10)$^d$              & $1.514^{+0.188}_{-0.193}$ & 7.677 $\pm$ 0.086 & L03b \\ 
{} UGC 4305 9       &  5.371 $\pm$ 0.379 & (10)$^d$              & $1.486^{+0.124}_{-0.126}$ & 7.699 $\pm$ 0.072 & L03b \\ 
{} UGC 4483         &  4.795 $\pm$ 0.052 & $  72^{  +93}_{ -72}$ & $1.657 \pm 0.026$         & 7.540 $\pm$ 0.012 & I94  \\ 
{} UGC 6456         &  5.918 $\pm$ 0.062 & $  29^{  +52}_{ -29}$ & $1.547 \pm 0.022$         & 7.696 $\pm$ 0.012 & I97  \\ 
{} UGC 6456 1       &  4.395 $\pm$ 0.399 & (10)$^d$              & $2.089^{+0.476}_{-0.436}$ & 7.355 $\pm$ 0.108 & L03b \\ 
{} UGC 6456 2       &  5.180 $\pm$ 0.372 & (10)$^d$              & $1.768^{+0.191}_{-0.188}$ & 7.519 $\pm$ 0.077 & L03b \\ 
{} UGC 9128         &  4.178 $\pm$ 0.166 & $ 198 \pm 16$         & $1.320^{+0.121}_{-0.127}$ & 7.745 $\pm$ 0.080 & L03b \\ 
{} UGC 9497 c       &  7.138 $\pm$ 0.101 & $<   77$              & $1.796 \pm 0.030$         & 7.608 $\pm$ 0.015 & G03b \\ 
{} UGC 9497 e       &  4.451 $\pm$ 0.307 & $<  453$              & $1.657^{+0.435}_{-0.440}$ & 7.524 $\pm$ 0.172 & G03b \\ 
{} UM 133           &  6.784 $\pm$ 0.112 & $<   26$              & $1.676 \pm 0.032$         & 7.692 $\pm$ 0.014 & I04b \\ 
{} UM 238           & 10.786 $\pm$ 0.219 & $ 288^{  +69}_{ -61}$ & $1.250 \pm 0.016$         & 8.177 $\pm$ 0.017 & I04b \\ 
{} UM 311           &  7.075 $\pm$ 0.077 & $  75^{  +40}_{ -36}$ & $0.977^{+0.037}_{-0.039}$ & 8.374 $\pm$ 0.044 & I98  \\ 
{} UM 396           &  9.391 $\pm$ 0.177 & $  37^{  +44}_{ -37}$ & $1.136 \pm 0.016$         & 8.238 $\pm$ 0.019 & I04b \\ 
{} UM 420           &  7.814 $\pm$ 0.182 & $<   79$              & $1.387^{+0.081}_{-0.083}$ & 7.941 $\pm$ 0.049 & I98  \\ 
{} UM 422           & 10.519 $\pm$ 0.200 & $<   57$              & $1.296 \pm 0.014$         & 8.121 $\pm$ 0.014 & I04b \\ 
{} UM 439           & 11.735 $\pm$ 0.226 & $ 177^{  +59}_{ -53}$ & $1.411 \pm 0.015$         & 8.069 $\pm$ 0.013 & I04b \\ 
{} UM 448           &  6.225 $\pm$ 0.067 & $ 138^{  +35}_{ -33}$ & $1.220^{+0.059}_{-0.061}$ & 8.018 $\pm$ 0.047 & I98  \\ 
{} UM 461           &  8.518 $\pm$ 0.200 & $ 203^{ +284}_{-194}$ & $1.615 \pm 0.042$         & 7.782 $\pm$ 0.027 & I98  \\ 
{} UM 462 SW        &  8.283 $\pm$ 0.052 & $<   10$              & $1.378 \pm 0.015$         & 7.960 $\pm$ 0.010 & I98  \\ 

\noalign{\smallskip}
\hline
\end{tabular}
\begin{list}{}{}
\item[$^{\mathrm{a}}$]
  Gas density of the S$^{+}$ regions
  in units of cm$^{-3}$.
\item[$^{\mathrm{b}}$]
  Gas temperature of the O$^{2+}$ regions
  in units of $10^4$K.
\item[$^{\mathrm{c}}$]
  References. ---
  G00:  Guseva et al. (2000),
  G03a: Guseva et al. (2003a),
  G03b: Guseva et al. (2003b),
  G03c: Guseva et al. (2003c),
  I94:  Izotov et al. (1994),
  I97:  Izotov et al. (1997),
  I98:  Izotov \& Thuan (1998),
  I04a: Izotov et al. (2004),
  I04b: Izotov \& Thuan (2004),
  I06:  Izotov et al. (2006a),
  K96:  Kobulnicky \& Skillman (1996),
  L03a: Lee et al. (2003a),
  L03b: Lee et al. (2003b),
  L04:  Lee et al. (2004),
  M04:  Melbourne et al. (2004),
  P92:  Pagel et al. (1992),
  P06:  Papaderos et al. (2006),
  T95:  Thuan et al. (1995),
  V03:  V\'{\i}lchez \& Iglesias-P\'{a}ramo (2003).
\item[$^{\mathrm{d}}$]
  Flux ratio of [S{\sc ii}] is not given in literature.
  We adopt $n_{\rm H}$(S$^+$) = 10 cm$^{-3}$ to calculate
  the oxygen abundance for these objects.
\end{list}
\end{table*}

\begin{table*}
\centering
\caption{Statistical properties of the samples}
\label{table:02a}
\begin{tabular}{l r c c c}
\hline\hline
\noalign{\smallskip}

Sample                & 
$N_{\rm obj}$         &
Median of 12+log(O/H) &
Mean of 12+log(O/H)   &
RMS of 12+log(O/H)   \\

\noalign{\smallskip}
\hline 
\noalign{\smallskip}

Sample A             &   139 & 8.010 & 8.003 & 0.233 \\
Sample B             &   120 & 7.936 & 7.858 & 0.303 \\
Sample C             & 48497 & 9.016 & 8.976 & 0.166 \\
\noalign{\smallskip}
Sample A+B           &   259 & 7.980 & 7.936 & 0.277 \\

\noalign{\smallskip}
\hline
\end{tabular}
\end{table*}

\begin{table*}
\vspace{12mm}
\centering
\caption{Means and RMSs of emission-line flux ratios of 
         the galaxies in the sample A+B$^{\mathrm{a}}$}
\label{table:03}
\begin{tabular}{l c c c c c c}
\hline\hline
\noalign{\smallskip}

Oxygen Abundance                                                    & 
log$R_{23}$                                                         &
log$\frac{F([{\rm NII}]\lambda6584) }{F({\rm H}\alpha)}$            &
log$\frac{F([{\rm OIII}]\lambda5007)}{F([{\rm NII}]\lambda6584)}$   &
log$\frac{F([{\rm NII}]\lambda6584) }{F([{\rm OII}]\lambda3727)}$   &
log$\frac{F([{\rm NII}]\lambda6584) }{F([{\rm SII}]\lambda6720)}$   &
log$\frac{F([{\rm OIII}]\lambda5007)}{F([{\rm OII}]\lambda3727)}$  \\

\noalign{\smallskip}
\hline 
\noalign{\smallskip}

$7.05 \! \leq \! 12 \! + \! {\rm log (O/H)} \! < \! 7.45$ & 
     0.601  & --2.452 &  2.415  & --1.657 & --0.668 &  0.920  \\
  & (0.108) & (0.238) & (0.196) & (0.269) & (0.178) & (0.247) \\
\noalign{\smallskip}
$7.45 \! \leq \! 12 \! + \! {\rm log (O/H)} \! < \! 7.55$ & 
     0.780  & --2.122 &  2.278  & --1.477 & --0.606 &  0.814  \\
  & (0.084) & (0.272) & (0.343) & (0.120) & (0.111) & (0.262) \\
\noalign{\smallskip}
$7.55 \! \leq \! 12 \! + \! {\rm log (O/H)} \! < \! 7.65$ & 
     0.809  & --1.844 &  2.006  & --1.376 & --0.566 &  0.652  \\
  & (0.068) & (0.195) & (0.274) & (0.183) & (0.197) & (0.356) \\
\noalign{\smallskip}
$7.65 \! \leq \! 12 \! + \! {\rm log (O/H)} \! < \! 7.75$ & 
     0.843  & --1.846 &  2.050  & --1.431 & --0.599 &  0.492  \\
  & (0.039) & (0.247) & (0.290) & (0.176) & (0.075) & (0.233) \\
\noalign{\smallskip}
$7.75 \! \leq \! 12 \! + \! {\rm log (O/H)} \! < \! 7.85$ & 
     0.909  & --1.887 &  2.162  & --1.458 & --0.621 &  0.719  \\
  & (0.039) & (0.205) & (0.260) & (0.129) & (0.112) & (0.233) \\
\noalign{\smallskip}
$7.85 \! \leq \! 12 \! + \! {\rm log (O/H)} \! < \! 7.95$ & 
     0.928  & --1.624 &  1.884  & --1.281 & --0.531 &  0.579  \\
  & (0.045) & (0.197) & (0.262) & (0.167) & (0.140) & (0.307) \\
\noalign{\smallskip}
$7.95 \! \leq \! 12 \! + \! {\rm log (O/H)} \! < \! 8.05$ & 
     0.941  & --1.578 &  1.858  & --1.262 & --0.484 &  0.571  \\
  & (0.042) & (0.213) & (0.280) & (0.179) & (0.118) & (0.218) \\
\noalign{\smallskip}
$8.05 \! \leq \! 12 \! + \! {\rm log (O/H)} \! < \! 8.15$ & 
     0.944  & --1.481 &  1.759  & --1.227 & --0.450 &  0.511  \\
  & (0.053) & (0.177) & (0.255) & (0.149) & (0.109) & (0.247) \\
\noalign{\smallskip}
$8.15 \! \leq \! 12 \! + \! {\rm log (O/H)} \! < \! 8.25$ & 
     0.938  & --1.375 &  1.631  & --1.119 & --0.348 &  0.514  \\
  & (0.044) & (0.142) & (0.185) & (0.114) & (0.120) & (0.168) \\
\noalign{\smallskip}
$8.25 \! \leq \! 12 \! + \! {\rm log (O/H)} \! < \! 8.35$ & 
     0.917  & --1.415 &  1.641  & --1.201 & --0.333 &  0.436  \\
  & (0.036) & (0.196) & (0.260) & (0.101) & (0.064) & (0.288) \\
\noalign{\smallskip}
$8.35 \! \leq \! 12 \! + \! {\rm log (O/H)} \! < \! 8.55$ & 
     0.892  & --1.124 &  1.259  & --1.084 & --0.173 &  0.258  \\
  & (0.034) & (0.176) & (0.221) & (0.109) & (0.122) & (0.087) \\

\noalign{\smallskip}
\hline
\end{tabular}
\begin{list}{}{}
\item[$^{\mathrm{a}}$]
  Mean and RMS of each emission-line flux ratio are given
  in the upper and lower rows. RMSs are given in parenthesis.
\end{list}
\end{table*}

\begin{table*}
\centering
\caption{Means and RMSs of emission-line flux ratios of 
         the galaxies in sample C$^{\mathrm{a}}$}
\label{table:04}
\begin{tabular}{l c c c c c c}
\hline\hline
\noalign{\smallskip}

Oxygen Abundance                                                    & 
log$R_{23}$                                                         &
log$\frac{F([{\rm NII}]\lambda6584) }{F({\rm H}\alpha)}$            &
log$\frac{F([{\rm OIII}]\lambda5007)}{F([{\rm NII}]\lambda6584)}$   &
log$\frac{F([{\rm NII}]\lambda6584) }{F([{\rm OII}]\lambda3727)}$   &
log$\frac{F([{\rm NII}]\lambda6584) }{F([{\rm SII}]\lambda6720)}$   &
log$\frac{F([{\rm OIII}]\lambda5007)}{F([{\rm OII}]\lambda3727)}$  \\

\noalign{\smallskip}
\hline 
\noalign{\smallskip}

$8.15 \! \leq \! 12 \! + \! {\rm log (O/H)} \! < \! 8.25$ & 
     0.835  & --1.138 &  1.102  & --1.196 & --0.532 & --0.089 \\ 
  & (0.043) & (0.081) & (0.171) & (0.077) & (0.042) & (0.140) \\
\noalign{\smallskip}
$8.25 \! \leq \! 12 \! + \! {\rm log (O/H)} \! < \! 8.35$ & 
     0.825  & --1.031 &  0.936  & --1.113 & --0.480 & --0.191 \\ 
  & (0.053) & (0.077) & (0.187) & (0.082) & (0.038) & (0.170) \\
\noalign{\smallskip}
$8.35 \! \leq \! 12 \! + \! {\rm log (O/H)} \! < \! 8.45$ & 
     0.817  & --0.934 &  0.768  & --1.056 & --0.444 & --0.283 \\ 
  & (0.060) & (0.088) & (0.199) & (0.084) & (0.033) & (0.174) \\
\noalign{\smallskip}
$8.45 \! \leq \! 12 \! + \! {\rm log (O/H)} \! < \! 8.55$ & 
     0.805  & --0.851 &  0.644  & --0.985 & --0.397 & --0.309 \\ 
  & (0.061) & (0.119) & (0.237) & (0.102) & (0.034) & (0.184) \\
\noalign{\smallskip}
$8.55 \! \leq \! 12 \! + \! {\rm log (O/H)} \! < \! 8.65$ & 
     0.772  & --0.812 &  0.572  & --0.910 & --0.334 & --0.329 \\ 
  & (0.052) & (0.101) & (0.216) & (0.091) & (0.047) & (0.170) \\
\noalign{\smallskip}
$8.65 \! \leq \! 12 \! + \! {\rm log (O/H)} \! < \! 8.75$ & 
     0.711  & --0.689 &  0.300  & --0.780 & --0.238 & --0.466 \\ 
  & (0.052) & (0.083) & (0.188) & (0.084) & (0.051) & (0.147) \\
\noalign{\smallskip}
$8.75 \! \leq \! 12 \! + \! {\rm log (O/H)} \! < \! 8.85$ & 
     0.637  & --0.596 &  0.060  & --0.642 & --0.146 & --0.573 \\ 
  & (0.052) & (0.068) & (0.160) & (0.075) & (0.051) & (0.135) \\
\noalign{\smallskip}
$8.85 \! \leq \! 12 \! + \! {\rm log (O/H)} \! < \! 8.95$ & 
     0.559  & --0.516 & --0.159 & --0.508 & --0.053 & --0.663 \\ 
  & (0.058) & (0.058) & (0.139) & (0.077) & (0.052) & (0.120) \\
\noalign{\smallskip}
$8.95 \! \leq \! 12 \! + \! {\rm log (O/H)} \! < \! 9.05$ & 
     0.454  & --0.447 & --0.383 & --0.345 &  0.054  & --0.731 \\ 
  & (0.071) & (0.056) & (0.123) & (0.088) & (0.057) & (0.108) \\
\noalign{\smallskip}
$9.05 \! \leq \! 12 \! + \! {\rm log (O/H)} \! < \! 9.15$ & 
     0.324  & --0.415 & --0.591 & --0.174 &  0.171  & --0.777 \\
  & (0.084) & (0.052) & (0.126) & (0.100) & (0.057) & (0.101) \\
\noalign{\smallskip}
$9.15 \! \leq \! 12 \! + \! {\rm log (O/H)} \! < \! 9.25$ & 
     0.168  & --0.398 & --0.706 &  0.016  &  0.286  & --0.758 \\
  & (0.074) & (0.056) & (0.192) & (0.085) & (0.044) & (0.118) \\

\noalign{\smallskip}
\hline
\end{tabular}
\begin{list}{}{}
\item[$^{\mathrm{a}}$]
  Mean and RMS of each emission-line flux ratio are given
  in the upper and lower rows. RMSs are given in parenthesis.
\end{list}
\end{table*}

\begin{table*}
\vspace{12mm}
\centering
\caption{Coefficients of the 
         best-fit polynomials for the observed relations between 
         the emission-line flux ratios and the oxygen abundance,
         where log$R$ = $a_0 + a_1 x + a_2 x^2 + a_3 x^3$
         [$x \equiv$ log($Z$/$Z_\odot$) $\equiv$ 12+log(O/H)--8.69].}
\label{table:05}
\begin{tabular}{l c c c c c c}
\hline\hline
\noalign{\smallskip}

Flux ratio (log$R$) & 
$a_0$ &
$a_1$ &
$a_2$ &
$a_3$ \\

\noalign{\smallskip}
\hline 
\noalign{\smallskip}

log $R_{23}$ &
   +7.1806E--1 & --6.9548E--1 & --6.2220E--1 & --6.3169E--2 \\
log [$F$([N{\sc ii}]$\lambda6584$)/$F$(H$\alpha$)] &
  --6.8307E--1 &  +8.9881E--1 & --5.2302E--1 & --2.2040E--1 \\ 
log [$F$([O{\sc iii}]$\lambda$5007)/$F$([N{\sc ii}]$\lambda$6584)] &
   +3.2921E--1 & --2.2578E+0  & --4.1699E--2 &  +3.7941E--1 \\
log [$F$([N{\sc ii}]$\lambda$6584)/$F$([O{\sc ii}]$\lambda$3727)] &
  --7.9322E--1 &  +1.1399E+0  &  +7.8929E--1 &  +2.7101E--1 \\
log [$F$([N{\sc ii}]$\lambda$6584)/$F$([S{\sc ii}]$\lambda$6720)] &
  --2.5214E--1 &  +7.4100E--1 &  +5.8181E--1 &  +1.7963E--1 \\
log [$F$([O{\sc iii}]$\lambda$5007)/$F$([O{\sc ii}]$\lambda$3727)] &
  --3.0777E--1 & --1.1210E+0  & --1.4359E--1 & --- \\
% log [$F$([Ne{\sc iii}]$\lambda$3869)/$F$([O{\sc ii}]$\lambda$3727)] &
%   --1.2547E+0  & --7.0929E--1 &  +3.0497E--1 &  +1.6784E--1 \\

\noalign{\smallskip}
\hline
\end{tabular}
\end{table*}

\clearpage

\begin{table*}
\centering
\caption{Coefficients of the 
         best-fit polynomials for the observed relations between 
         the emission-line flux ratios and the oxygen abundance,
         where log$R$ = $b_0 + b_1 y + b_2 y^2 + b_3 y^3$
         [$y \equiv$ 12+log(O/H)].}
\label{table:06}
\begin{tabular}{l c c c c c c}
\hline\hline
\noalign{\smallskip}

Flux ratio (log$R$) & 
$b_0$ &
$b_1$ &
$b_2$ &
$b_3$ \\

\noalign{\smallskip}
\hline 
\noalign{\smallskip}

log $R_{23}$ &
   +1.2299E+0 & --4.1926E+0 &  +1.0246E+0 & --6.3169E--2 \\
log [$F$([N{\sc ii}]$\lambda6584$)/$F$(H$\alpha$)] &
   +9.6641E+1 & --3.9941E+1 &  +5.2227E+0 & --2.2040E--1 \\ 
log [$F$([O{\sc iii}]$\lambda$5007)/$F$([N{\sc ii}]$\lambda$6584)] &
  --2.3218E+2 &  +8.4423E+1 & --9.9330E+0 &  +3.7941E--1 \\
log [$F$([N{\sc ii}]$\lambda$6584)/$F$([O{\sc ii}]$\lambda$3727)] &
  --1.2894E+2 &  +4.8818E+1 & --6.2759E+0 &  +2.7101E--1 \\
log [$F$([N{\sc ii}]$\lambda$6584)/$F$([S{\sc ii}]$\lambda$6720)] &
  --8.0632E+1 &  +3.1323E+1 & --4.1010E+0 &  +1.7963E--1 \\
log [$F$([O{\sc iii}]$\lambda$5007)/$F$([O{\sc ii}]$\lambda$3727)] &
  --1.4089E+0 &  +1.3745E+0 & --1.4359E--1& --- \\
% log [$F$([Ne{\sc iii}]$\lambda$3869)/$F$([O{\sc ii}]$\lambda$3727)] &
%   --8.2202E+1 &  +3.2014E+1 & --4.0706E+0 &  +1.6784E--1 \\

\noalign{\smallskip}
\hline
\end{tabular}
\end{table*}

\begin{table*}
\vspace{12mm}
\centering
\caption{Means and RMSs of additional emission-line flux ratios of 
         the galaxies in the sample A+B$^{\mathrm{a}}$}
\label{table:07}
\begin{tabular}{l c c c c}
\hline\hline
\noalign{\smallskip}

Oxygen Abundance                                                    & 
log$\frac{F([{\rm NeIII}]\lambda3869) }{F([{\rm OII}]\lambda3727)}$ &
log$\frac{F({\rm H}\alpha+{\rm [NII]}\lambda\lambda6548,6584)}
         {F([{\rm SII}]\lambda6720)}$                               &
log$\frac{F([{\rm OIII}]\lambda5007)}{F({\rm H}\beta)}$             &
log$\frac{F([{\rm OII}]\lambda3727)}{F({\rm H}\beta)}$             \\

\noalign{\smallskip}
\hline 
\noalign{\smallskip}

$7.05 \! \leq \! 12 \! + \! {\rm log (O/H)} \! < \! 7.45$ & 
    --0.092 &  1.737  &  0.421  & --0.486  \\
  & (0.144) & (0.156) & (0.134) & (0.186)  \\
\noalign{\smallskip}
$7.45 \! \leq \! 12 \! + \! {\rm log (O/H)} \! < \! 7.55$ & 
    --0.285 &  1.481  &  0.594  & --0.210  \\
  & (0.255) & (0.209) & (0.100) & (0.161)  \\
\noalign{\smallskip}
$7.55 \! \leq \! 12 \! + \! {\rm log (O/H)} \! < \! 7.65$ & 
    --0.426 &  1.254  &  0.600  & --0.050  \\
  & (0.368) & (0.256) & (0.124) & (0.226)  \\
\noalign{\smallskip}
$7.65 \! \leq \! 12 \! + \! {\rm log (O/H)} \! < \! 7.75$ & 
    --0.545 &  1.188  &  0.653  &  0.127   \\
  & (0.202) & (0.175) & (0.062) & (0.193)  \\
\noalign{\smallskip}
$7.75 \! \leq \! 12 \! + \! {\rm log (O/H)} \! < \! 7.85$ & 
    --0.374 &  1.217  &  0.724  &  0.003   \\
  & (0.219) & (0.177) & (0.072) & (0.171)  \\
\noalign{\smallskip}
$7.85 \! \leq \! 12 \! + \! {\rm log (O/H)} \! < \! 7.95$ & 
    --0.504 &  1.038  &  0.708  &  0.140   \\
  & (0.287) & (0.171) & (0.082) & (0.216)  \\
\noalign{\smallskip}
$7.95 \! \leq \! 12 \! + \! {\rm log (O/H)} \! < \! 8.05$ & 
    --0.510 &  1.036  &  0.731  &  0.163   \\
  & (0.222) & (0.149) & (0.082) & (0.144)  \\
\noalign{\smallskip}
$8.05 \! \leq \! 12 \! + \! {\rm log (O/H)} \! < \! 8.15$ & 
    --0.592 &  0.992  &  0.732  &  0.216   \\
  & (0.228) & (0.150) & (0.092) & (0.163)  \\
\noalign{\smallskip}
$8.15 \! \leq \! 12 \! + \! {\rm log (O/H)} \! < \! 8.25$ & 
    --0.603 &  0.968  &  0.712  &  0.215   \\
  & (0.150) & (0.101) & (0.069) & (0.103)  \\
\noalign{\smallskip}
$8.25 \! \leq \! 12 \! + \! {\rm log (O/H)} \! < \! 8.35$ & 
    --0.694 &  1.038  &  0.684  &  0.244   \\
  & (0.346) & (0.197) & (0.086) & (0.188)  \\
\noalign{\smallskip}
$8.35 \! \leq \! 12 \! + \! {\rm log (O/H)} \! < \! 8.55$ & 
    --0.900 &  0.932  &  0.590  &  0.361   \\
  & (0.095) & (0.082) & (0.060) & (0.074)  \\

\noalign{\smallskip}
\hline
\end{tabular}
\begin{list}{}{}
\item[$^{\mathrm{a}}$]
  Mean and RMS of each emission-line flux ratio are given
  in the upper and lower rows. RMSs are given in parenthesis.
\end{list}
\end{table*}

\begin{table*}
\centering
\caption{Means and RMSs of additional emission-line flux ratios of 
         the galaxies in sample C$^{\mathrm{a}}$}
\label{table:08}
\begin{tabular}{l c c c c}
\hline\hline
\noalign{\smallskip}

Oxygen Abundance                                                    &
log$\frac{F([{\rm NeIII}]\lambda3869) }{F([{\rm OII}]\lambda3727)}$ &
log$\frac{F({\rm H}\alpha+{\rm [NII]}\lambda\lambda6548,6584)}
         {F([{\rm SII}]\lambda6720)}$                               &
log$\frac{F([{\rm OIII}]\lambda5007)}{F({\rm H}\beta)}$             &
log$\frac{F([{\rm OII}]\lambda3727)}{F({\rm H}\beta)}$             \\

\noalign{\smallskip}
\hline 
\noalign{\smallskip}

$8.15 \! \leq \! 12 \! + \! {\rm log (O/H)} \! < \! 8.25$ & 
    --1.060 &  0.637  &  0.405  &  0.512   \\ 
  & (0.112) & (0.081) & (0.102) & (0.062)  \\
\noalign{\smallskip}
$8.25 \! \leq \! 12 \! + \! {\rm log (O/H)} \! < \! 8.35$ & 
    --1.084 &  0.591  &  0.357  &  0.549   \\ 
  & (0.117) & (0.089) & (0.119) & (0.079)  \\
\noalign{\smallskip}
$8.35 \! \leq \! 12 \! + \! {\rm log (O/H)} \! < \! 8.45$ & 
    --1.132 &  0.540  &  0.281  &  0.580   \\ 
  & (0.111) & (0.098) & (0.123) & (0.091)  \\
\noalign{\smallskip}
$8.45 \! \leq \! 12 \! + \! {\rm log (O/H)} \! < \! 8.55$ & 
    --1.113 &  0.518  &  0.250  &  0.578   \\ 
  & (0.109) & (0.105) & (0.134) & (0.086)  \\
\noalign{\smallskip}
$8.55 \! \leq \! 12 \! + \! {\rm log (O/H)} \! < \! 8.65$ & 
    --1.190 &  0.545  &  0.219  &  0.555   \\ 
  & (0.121) & (0.087) & (0.129) & (0.075)  \\
\noalign{\smallskip}
$8.65 \! \leq \! 12 \! + \! {\rm log (O/H)} \! < \! 8.75$ & 
    --1.301 &  0.538  &  0.070  &  0.544   \\ 
  & (0.121) & (0.068) & (0.124) & (0.062)  \\
\noalign{\smallskip}
$8.75 \! \leq \! 12 \! + \! {\rm log (O/H)} \! < \! 8.85$ & 
    --1.339 &  0.553  & --0.077 &  0.501   \\ 
  & (0.121) & (0.057) & (0.120) & (0.057)  \\
\noalign{\smallskip}
$8.85 \! \leq \! 12 \! + \! {\rm log (O/H)} \! < \! 8.95$ & 
    --1.400 &  0.583  & --0.218 &  0.446   \\ 
  & (0.123) & (0.048) & (0.116) & (0.060)  \\
\noalign{\smallskip}
$8.95 \! \leq \! 12 \! + \! {\rm log (O/H)} \! < \! 9.05$ & 
    --1.371 &  0.639  & --0.369 &  0.352   \\ 
  & (0.266) & (0.043) & (0.116) & (0.074)  \\
\noalign{\smallskip}
$9.05 \! \leq \! 12 \! + \! {\rm log (O/H)} \! < \! 9.15$ & 
      ---   &  0.727  & --0.530 &  0.228   \\
  & ( --- ) & (0.047) & (0.119) & (0.089)  \\
\noalign{\smallskip}
$9.15 \! \leq \! 12 \! + \! {\rm log (O/H)} \! < \! 9.25$ & 
      ---   &  0.823  & --0.611 &  0.078   \\
  & ( --- ) & (0.043) & (0.177) & (0.085)  \\

\noalign{\smallskip}
\hline
\end{tabular}
\begin{list}{}{}
\item[$^{\mathrm{a}}$]
  Mean and RMS of each emission-line flux ratio are given
  in the upper and lower rows. RMSs are given in parenthesis.
\end{list}
\end{table*}

\begin{table*}
\vspace{12mm}
\centering
\caption{Coefficients of the 
         best-fit polynomials for the observed relations between 
         the additional emission-line flux ratios and the oxygen abundance,
         where log$R$ = $a_0 + a_1 x + a_2 x^2 + a_3 x^3$
         [$x \equiv$ log($Z$/$Z_\odot$) $\equiv$ 12+log(O/H)--8.69].}
\label{table:09}
\begin{tabular}{l c c c c c c}
\hline\hline
\noalign{\smallskip}

Flux ratio (log$R$) & 
$a_0$ &
$a_1$ &
$a_2$ &
$a_3$ \\

\noalign{\smallskip}
\hline 
\noalign{\smallskip}

log [$F$([Ne{\sc iii}]$\lambda$3869)/$F$([O{\sc ii}]$\lambda$3727)] &
  --1.2547E+0  & --7.0929E--1 &  +3.0497E--1 &  +1.6784E--1 \\
log [$F$(H$\alpha$+[N{\sc ii}]$\lambda\lambda$6548,6584)/$F$([S{\sc ii}]$\lambda$6720)] &
   +5.6097E--1 & --7.9971E--2 &  +9.8562E--1 &  +3.4069E--1 \\
log [$F$([O{\sc iii}]$\lambda$5007)/$F$(H$\beta$)] &
   +1.6366E--1 & --1.3785E+0  & --8.4778E--1 &  +9.1853E--3 \\
log [$F$([O{\sc ii}]$\lambda$3727)/$F$(H$\beta$)] &
   +5.3481E--1 & --2.0792E--1 & --1.1353E+0  & --3.5951E--1 \\

\noalign{\smallskip}
\hline
\end{tabular}
\end{table*}

\begin{table*}
\vspace{12mm}
\centering
\caption{Coefficients of the 
         best-fit polynomials for the observed relations between 
         the additional emission-line flux ratios and the oxygen abundance,
         where log$R$ = $b_0 + b_1 y + b_2 y^2 + b_3 y^3$
         [$y \equiv$ 12+log(O/H)].}
\label{table:10}
\begin{tabular}{l c c c c c c}
\hline\hline
\noalign{\smallskip}

Flux ratio (log$R$) & 
$b_0$ &
$b_1$ &
$b_2$ &
$b_3$ \\

\noalign{\smallskip}
\hline 
\noalign{\smallskip}

log [$F$([Ne{\sc iii}]$\lambda$3869)/$F$([O{\sc ii}]$\lambda$3727)] &
  --8.2202E+1 &  +3.2014E+1 & --4.0706E+0 &  +1.6784E--1 \\
log [$F$(H$\alpha$+[N{\sc ii}]$\lambda\lambda$6548,6584)/$F$([S{\sc ii}]$\lambda$6720)] &
  --1.4789E+2 &  +5.9974E+1 & --7.8963E+0 &  +3.4069E--1 \\
log [$F$([O{\sc iii}]$\lambda$5007)/$F$(H$\beta$)] &
  --5.7906E+1 &  +1.5437E+1 & --1.0872E+0 &  +9.1853E--3 \\
log [$F$([O{\sc ii}]$\lambda$3727)/$F$(H$\beta$)] &
   +1.5253E+2 & --6.1922E+1 &  +8.2370E+0 & --3.5951E--1 \\

\noalign{\smallskip}
\hline
\end{tabular}
\end{table*}

\end{document}